\pdfoutput=1 
\documentclass[acmsmall]{acmart}

\usepackage{CJKutf8}
\usepackage{url}
\usepackage{booktabs}
\usepackage{enumitem}
\usepackage{makecell}
\usepackage{hyperref}
\usepackage{xcolor}
\usepackage{graphicx}
\usepackage[export]{adjustbox}
\usepackage{xspace}
\usepackage{multirow}

\newcommand{\fix}[1]{}

\newcommand{\umlgraph}{\emph{UMLGraph}\xspace}
\newcommand{\mrgnn}{MRGNN\xspace}

\newcommand{\codenet}{CoCoNet\xspace}

\newcommand{\codeuml}{CoCoSUM\xspace}
\newcommand{\codeumlh}{CoCoSUM$_h$\xspace}

\newcommand{\codeumlo}{CoCoSUM$_u$\xspace}
\newcommand{\bertatt}{CoCoSUM$_c$\xspace}
\newcommand{\astattgru}{Ast-attendgru\xspace}
\newcommand{\attgru}{Attendgru\xspace}
\newcommand{\neuralcs}{NeuralCodeSum\xspace}
\newcommand{\retocom}{Re2Com\xspace}
\newcommand{\hdeepcom}{H-Deepcom\xspace}
\newcommand{\codenn}{Code-NN\xspace}
\newcommand{\codebert}{CodeBert\xspace}
\newcommand{\hybriddrl}{Hybrid-DRL\xspace}
\newcommand{\codetoseq}{Code2seq\xspace}
\newcommand{\astnn}{ASTNN\xspace}
\newcommand{\rencos}{Rencos\xspace}

\newcommand{\transformer}{Transformer\xspace}

\newcommand{\Fig}{Figure\xspace}
\newcommand{\Sec}{Section\xspace}
\newcommand{\Tab}{Table\xspace}
\newcommand{\Eq}{Equation\xspace}

\usepackage{listings}

\definecolor{light-gray}{gray}{0.95}
\usepackage{caption}
\usepackage{fancybox}
\usepackage[most]{tcolorbox}


\usepackage[caption=false,font=footnotesize]{subfig}

\AtBeginDocument{%
  \providecommand\BibTeX{{%
    \normalfont B\kern-0.5em{\scshape i\kern-0.25em b}\kern-0.8em\TeX}}}

\setcopyright{acmcopyright}

\acmJournal{JACM}

\begin{document}

\pagestyle{plain}
\title{\codeuml: Contextual Code Summarization with Multi-Relational Graph Neural Network}

\author{Yanlin Wang}
\affiliation{%
  \institution{Microsoft Research Asia}
  \city{Beijing}
  \country{China}}
\email{yanlwang@microsoft.com}

\author{Ensheng Shi}
\affiliation{%
  \institution{Xi'an Jiaotong University}
  \city{Xi'an}
  \country{China}}
\email{s1530129650@stu.xjtu.edu.cn}

\author{Lun Du}
\affiliation{%
  \institution{Microsoft Research Asia}
  \city{Beijing}
  \country{China}}
\email{lun.du@microsoft.com}

\author{Xiaodi Yang}
\affiliation{%
  \institution{The University of Hong Kong}
  \city{Hong Kong}
  \country{China}}
\email{s1530129650@stu.xjtu.edu.cn}

\author{Yuxuan Hu}
\affiliation{%
  \institution{Beijing University of Technology}
  \city{Beijing}
  \country{China}}
\email{s1530129650@stu.xjtu.edu.cn}

\author{Shi han}
\affiliation{%
  \institution{Microsoft Research Asia}
  \city{Beijing}
  \country{China}}
\email{shihan@microsoft.com}

\author{Hongyu Zhang}
\affiliation{%
  \institution{The University of Newcastle}
  \city{}
  \country{Australia}}
\email{hongyu.zhang@newcastle.edu.au}

\author{Dongmei Zhang}
\affiliation{%
  \institution{Microsoft Research Asia}
  \city{Beijing}
  \country{China}}
\email{dongmeiz@microsoft.com}


\begin{abstract}


Source code summaries are short natural language descriptions of code snippets
that help developers better understand and maintain source code. 
There has been a surge of work on automatic code summarization to reduce the burden of writing summaries manually. However, most contemporary approaches mainly leverage the information within the boundary of the method being summarized (i.e., local context), and ignore the broader context that could assist with code summarization. This paper explores two global contexts, namely intra-class and inter-class contexts, and proposes the model \codeuml{}: Contextual Code Summarization
with Multi-Relational Graph Neural Networks. 
\codeuml{} first incorporates class names as the intra-class context
to generate the \emph{class semantic embeddings}. Then, relevant Unified Modeling Language (UML) class diagrams are
extracted as inter-class context and are encoded into the \emph{class
relational embeddings} using a novel Multi-Relational Graph Neural Network
(\mrgnn{}). 
Class semantic embeddings and class relational embeddings, together with the outputs from code token encoder and AST encoder, are passed to a decoder armed with a two-level attention mechanism to generate high-quality, context-aware code summaries. We conduct extensive experiments to evaluate our approach and compare it with other automatic code summarization models. The experimental results show that \codeuml{} is effective and outperforms state-of-the-art methods. 

\end{abstract}



\maketitle

\section{Introduction}\label{sec:introduction}
Code summaries (i.e., source code comments) are short natural language descriptions of source code. As it often takes developers much time to read and understand other people's code, good code summaries can facilitate program comprehension by helping developers quickly understand what a piece of code does. Therefore, code summaries are important for software development and maintenance. 
However,  many developers are  not used to write  comments  and  many  comments  are  mismatched, missing, or outdated.  For example, Spinellis ~\cite{spinellis2010code} found that, in third-party projects ported to the FreeBSD platform, the number of comments per 100 lines varies substantially and even worse, the code of 307 projects has less than one comment per 100 lines. It is thus desirable that source code can be automatically summarized in natural language.

There is a surge of work on automatic code summarization for better program comprehension~\cite{SridharaHMPV10,HaiducAM10ICSE,IyerKCZ16,WanZYXY0Y18,hu2019deep,LeClairJM19,AhmadCRC20,leclair2020gnn,fernandes2018structured,fowkes2016tassal,Movshovitz-AttiasC13,AllamanisBBS15,AllamanisPS16,haije2016automatic,HuLXLJ18,wei2019code,haque2020improved,WeiLLXJ20,zhangretrieval20}.
Rule-based approaches rely on predefined rules or templates~\cite{SridharaHMPV10} and are thus limited in the types of summaries that can be generated. Information Retrieval (IR) based approaches use similar code snippets to select or synthesize a comment~\cite{HaiducAM10ICSE}, thus lacking the capability of learning semantic meanings from existing code. 
With the accumulation of publicly available source code, data-driven approaches based on deep learning techniques have largely overtaken traditional methods. Recently, inspired by the success of Neural Machine Translation (NMT)~\cite{ChoMGBBSB14}, some researchers have applied NMT models to the code summarization task~\cite{IyerKCZ16, WanZYXY0Y18, hu2019deep, LeClairJM19}.


We have noticed that most previous code summarization techniques~\cite{Movshovitz-AttiasC13,AllamanisBBS15,AllamanisPS16,IyerKCZ16,haije2016automatic,HuLXLJ18,WanZYXY0Y18,hu2019deep,LeClairJM19,WeiLLXJ20,zhangretrieval20,AhmadCRC20} only leverage the information (code sequence, AST, etc) within the boundary of the source code method being summarized, which means that only \emph{local} context has been used. 
Global contexts, such as 
the enclosing class (i.e., the class enclosing the target method)
and class dependencies, provide important information that local context cannot cover. 
However, global contexts have not received much attention in source code summarization.
From the perspective of Object-Oriented Programming (OOP), a method typically defines an operation conducted by its enclosing class, applying to the class itself or against other classes. The operation's initiator/receptor classes (e.g., keywords in class name) and their interactions (e.g., class-class level dependence) may be mentioned in a good summary, which is typically outside of the local context of a method. For example, the word \emph{resource} in the following method summary comes from its enclosing class name \texttt{Resource}:


\begin{tcolorbox}[colback=white,colframe=yellow!50!black,boxrule=0.2mm,bottom = 0pt]
\begin{lstlisting}[language=Java,escapechar=@,linewidth=0.99\columnwidth,xleftmargin=-12pt,frame=single,framesep=0mm,backgroundcolor=\color{white},tabsize=1]
class @\colorbox{yellow}{Resource}@{
    ...
    // check whether a @\colorbox{yellow}{resource}@ is occupied
    boolean isOccupied() { return ownerId >= 0; }
}
\end{lstlisting}
\end{tcolorbox}
\noindent Taking the following code snippet as another example, the word
\emph{compiler} in the method summary comes from class \texttt{Compiler}
which the enclosing class \texttt{Parser} \emph{extends}: 


\begin{tcolorbox}[colback=white,colframe=yellow!50!black,boxrule=0.2mm,bottom = 0pt]
\begin{lstlisting}[language=Java,escapechar=@,linewidth=0.99\columnwidth,xleftmargin=-12pt,frame=single,framesep=0mm,backgroundcolor=\color{white},tabsize=1]
class Parser extends @\colorbox{yellow}{Compiler}@ {
    ...
    // add source code to the @\colorbox{yellow}{compiler}@
    void addSource(String className,String sourceCode) {
        sourceCodes.put(className, new SourceCode(className, sourceCode)); }
}
\end{lstlisting}
\end{tcolorbox}
\noindent Based on such an insight, we argue that leveraging broader context of a source code method
would help better summarize the method.


In this paper, we propose \textbf{\codeuml}: \textbf{Co}ntextual \textbf{Co}de \textbf{SU}mmarization with \textbf{M}ulti-Relational Graph Neural Networks. 
It addresses the aforementioned limitations of existing work by exploring two global contexts: intra-class level context and inter-class level context. For the intra-class context, we aim to capture the $\langle$method, class$\rangle$ relationship by using class name information to enrich the generated code summaries. For the inter-class context, we choose the $\langle$class, class$\rangle$ relationship as the target, which can be represented as a Unified Modeling Language (UML) class diagram.
Nodes in a UML class diagram represent classes/interfaces and edges reflect the class-class relationship such as association, realization, dependency, etc. 


In our design of \codeuml{}, in addition to the conventional code token
encoder and AST encoder as in other
approaches~\cite{HuLXLJ18,WanZYXY0Y18,hu2019deep,LeClairJM19}, we incorporate
intra-class context by using the class name representation learned from a 
Transformer-based sentence embedding model~\cite{cer2018universal}. This is
used as an additional encoder to the prevalent two-encoder architecture 
(i.e., code token encoder and AST encoder). Moreover, we extend the idea of
graph neural network GAT~\cite{VelickovicCCRLB18} and design a novel
Multi-Relational Graph Neural Network (MRGNN) with attention mechanism to
encode the inter-class context represented by a UML class diagram. 


We conduct extensive experiments on the public dataset CodeSearchNet~\cite{abs-1909-09436} and our newly collected dataset \codenet. The results demonstrate that \codeuml{} outperforms the
state-of-the-art work with respect to four widely-used metrics (BLEU, ROUGE-L,
METEOR, and CIDER). This is because incorporating intra-class and inter-class
contexts and combining them with the attention mechanism improve the quality
of the generated code summaries. We also perform an extensive ablation study
to verify the usefulness of each major component of \codeuml{}. Furthermore,
we apply the global contexts to existing representative models and the
improved results of these models confirm the generality of the proposed approach. Finally, we
perform a human evaluation 
and the results further confirm the effectiveness of our approach.


We summarize our main contributions as follows:
\begin{itemize}

\item To our best knowledge, we are the first to model the two global contexts in automatic code summarization: class names as intra-class context and relationships among classes as inter-class context.

\item We propose the Multi-Relational Graph Neural Network (\mrgnn{}) to model
the UML diagrams of source code so that inter-class context can be captured. 

\item We design a novel end-to-end automatic code summarization framework
\codeuml{}, which incorporates two novel components: (1) a Transformer-based
sentence embedding model for encoding class names and (2) an MRGNN for modeling
UML class diagrams. 

\item Our extensive experiments demonstrate that \codeuml{}
outperforms state-of-the-art code summarization methods.
Furthermore,  the performance of existing code summarization models is also improved after applying the global contexts adopted in \codeuml{}.
\end{itemize}

The remainder of this paper is organized as follows: \Sec~\ref{sec:bg}
presents the background. \Sec~\ref{sec:model} introduces the design of
\codeuml{}. We present details about our experiment settings in
\Sec~\ref{sec:exp}. We compare \codeuml{} with state-of-the-art works, conduct ablation study, evaluate the stability and generality of \codeuml and conduct the human evaluation in
\Sec~\ref{sec:eval}. We discuss the threats to validity in \Sec~\ref{sec:threats}. \Sec~\ref{sec:rw} describes the related work. Finally, we conclude our work in \Sec~\ref{sec:con}. 
\section{Background}\label{sec:bg}
In this section, we briefly illustrate some background knowledge.

\subsection{UML Class Diagrams}


\begin{figure}[t]
\centering
\includegraphics[width=0.9\columnwidth]{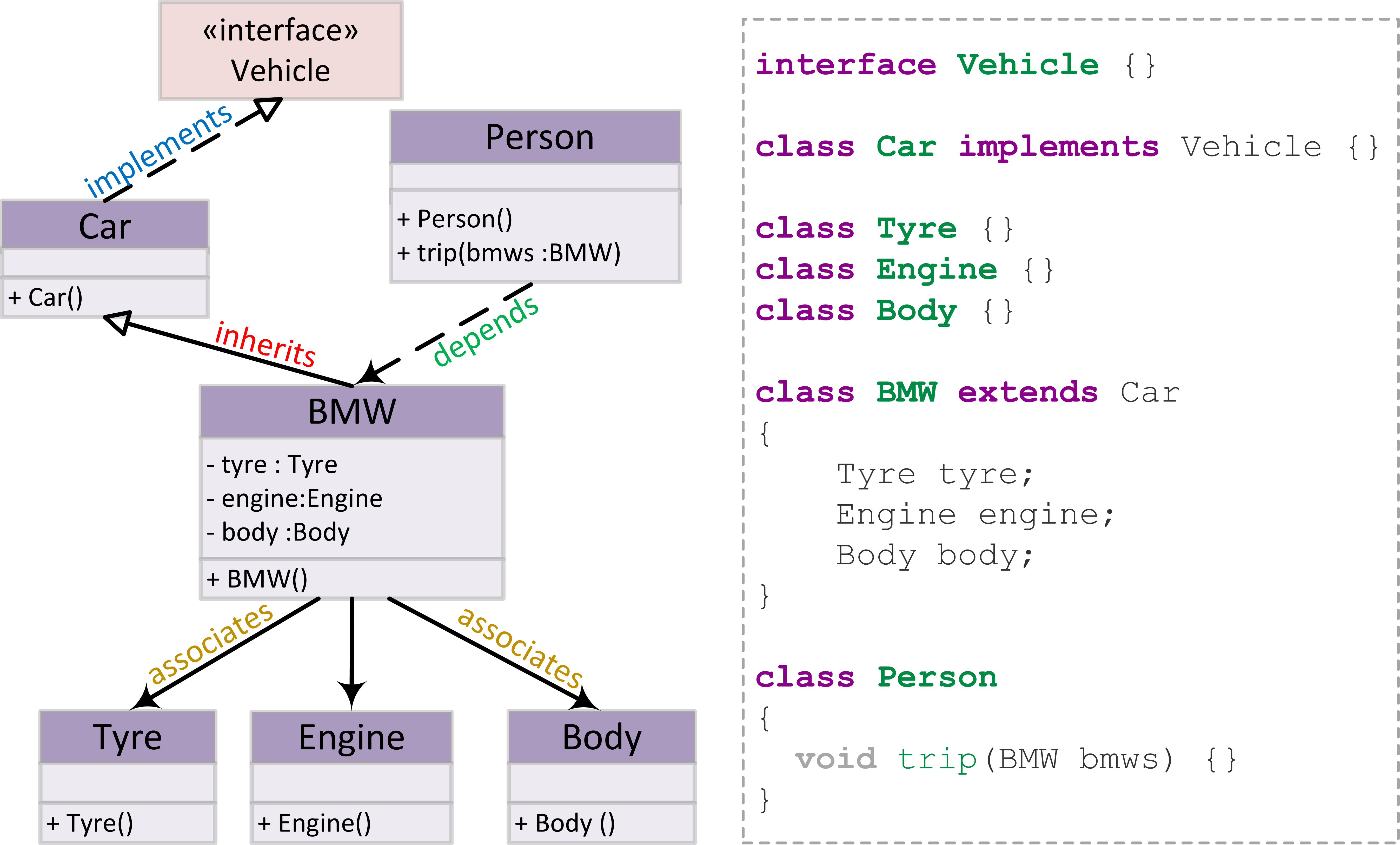}
\vspace{6pt}
\caption{UML class diagram example. Left is the UML class diagram and right is the corresponding code.}
\label{fig:uml_ex}
\vspace{5pt}
\end{figure}


The Unified Modeling Language (UML)~\cite{Booch99uml, Rumbaugh99uml} is a
widely-used language for specifying, visualizing, and documenting the
artifacts of a software-intensive system. 
UML class diagram is a type of static structure diagram that describes the
structure of a system by showing the system's classes and their attributes,
operations (or methods), and relationships. 
In this paper we extract four most common types of relationships expressed by
a UML class diagram: generalization, realization, dependency, and association.
These relationships are also supported by commonly-used UML tools such as
\umlgraph\footnote{\url{https://www.spinellis.gr/umlgraph/index.html}}.

\Fig~\ref{fig:uml_ex} shows a basic UML class diagram example. The class
\texttt{Car} implements the interface \texttt{Vehicle}. The dashed line with a
hollow arrow connecting class \texttt{Car} and interface \texttt{Vehicle}
indicates the `realization' relationship. The class \texttt{BMW} inherits from
the more general class \texttt{Car}. The `generalization' relationship is
represented by a solid line with a hollow arrow. Further, a \texttt{BMW}
class is associated with its children classes \texttt{Type}, \texttt{Engine}
and \texttt{Body}. The `association' relationship is represented by a solid
line with a solid arrow. The \texttt{Person} class depends on
\texttt{BMW} class for its method \texttt{trip}. The `dependency' relationship
is represented by a dashed line with a solid arrow.

\subsection{Seq2Seq Models}

Seq2Seq~\cite{SutskeverVL14} is a family of machine learning models, which
aims at transforming a sequence into another sequence. Typical applications
are machine translation~\cite{ChoMGBBSB14}, text
summarization~\cite{RushCW15}, image captioning~\cite{VinyalsTBE15}, etc. 
Seq2Seq approaches are based on an encoder-decoder architecture, with an
encoder mapping a source sequence to a latent vector and a decoder translating
the latent vector into a target sequence. Recurrent Neural Networks
(RNNs)~\cite{rumelhart1986learning}, Long Short-term Memory Networks
(LSTMs)~\cite{HochreiterS97}, Gated Recurrent Units (GRUs)~\cite{ChoMGBBSB14},
Transformers~\cite{VaswaniSPUJGKP17} or other deep neural
networks~\cite{YoungHPC18}, which are capable of modeling sequential data, 
are widely used by the encoder and the decoder in such a design. Furthermore,
there are some techniques that are commonly used in the encoder-decoder
architecture, including attention mechanism~\cite{VaswaniSPUJGKP17}, teacher
forcing~\cite{WilliamsZ89}, etc. 
Attention mechanism endows the decoder with the ability to process the input
sequence selectively. Teacher forcing uses the ground-truth sequences to guide
the generation in the training phase. 


\subsection{Graph Neural Networks}

The recent advances of deep learning techniques have facilitated many machine
learning tasks like object detection, machine translation, and speech
recognition~\cite{WuPCLZY20}. The key to such a technology evolution is 
deep neural networks such as Convolutional Neural Networks
(CNNs)~\cite{lecun2015deep}, which are able to automatically learn the data
features instead of heavily relying on the handcrafted features.

Unlike the data that are originally organized in euclidean space (e.g., images
on 2D grid and texts in 1D sequence), where CNNs and RNNs can effectively
capture hidden patterns, graph data is irregular and each node may have a
different number of neighbors. The irregularity of graph data makes some
important operations (e.g., convolution), which are easy to compute in
euclidean space, difficult for graph data~\cite{WuPCLZY20}. The attempt to
extend deep neural networks to graph data has spawned Graph Neural Networks
(GNNs)~\cite{abs-1812-08434,WuPCLZY20}.

The first motivation of GNNs roots in CNNs, since CNNs have a strong ability
to learn local stationary structures via localized convolution filter and
compose them to form multi-scale hierarchical patterns. Many efforts have been
devoted to defining the convolution operation for graph data. This branch of
GNNs is therefore called Graph Convolutional Networks (GCNs) and can be
divided into spectral methods (defining convolution in spectral domain) and
spatial methods (defining convolution in vertex domain).

The prevalence of graph data in real-world makes GNNs applicable to many
learning-based systems. Due to the strong ability to learn from graph data,
GNNs have achieved great successes in many tasks such as learning chemistry rule
and phenomenon~\cite{Hu20}, recommender systems~\cite{WuZGHWGC19}, document
classification~\cite{KipfW17}, traffic prediction~\cite{ZhangSXMKY18}, etc.

\section{\codeuml{} Model}~\label{sec:model}




\begin{figure}[t]
\includegraphics[width=0.7\columnwidth]{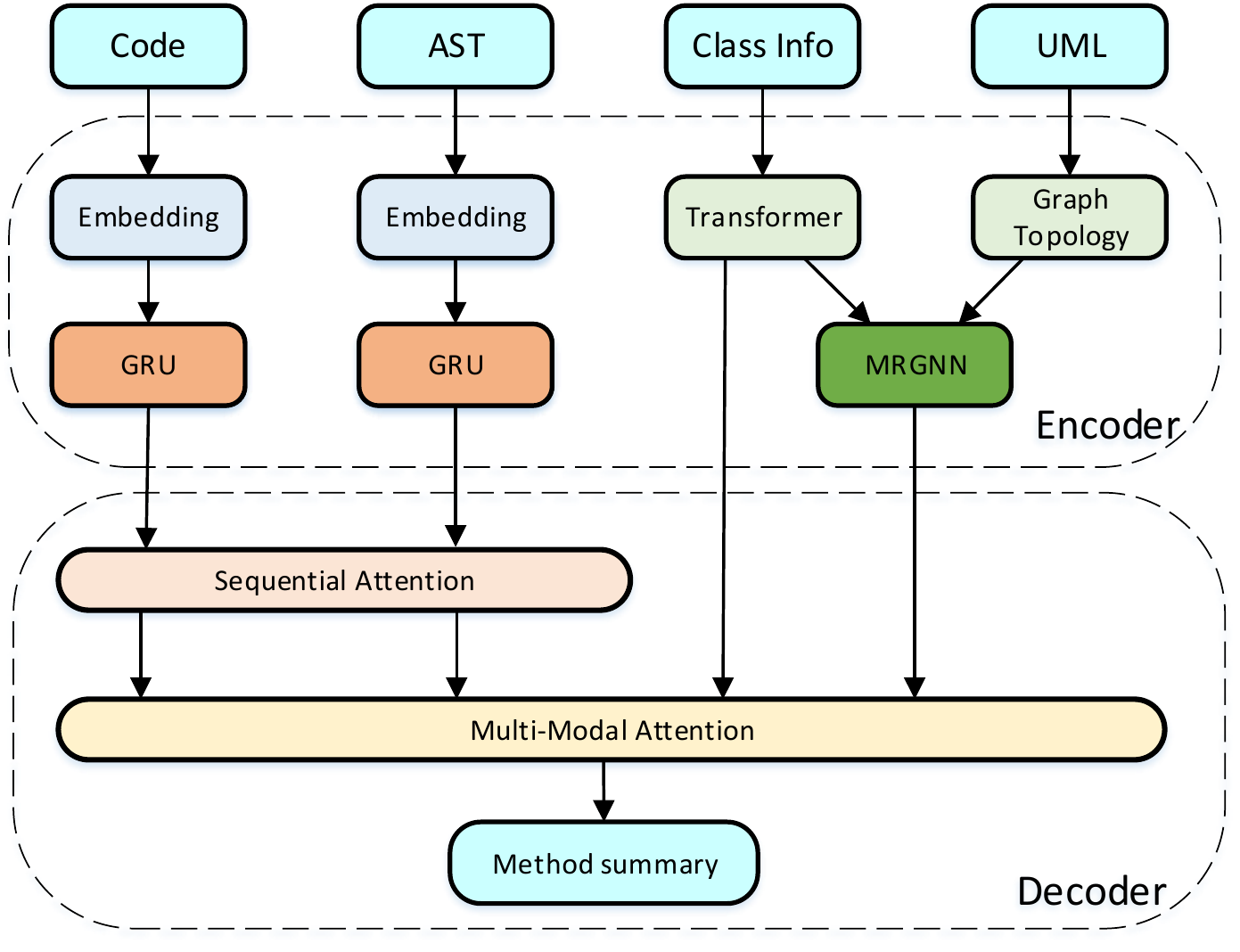}
\vspace{5pt}
\caption{The overview of \codeuml} 
\label{fig:overview}
\vspace{10pt}
\end{figure}


In this section, we present our model \codeuml{}. As shown in \Fig~\ref{fig:overview}, the architecture
follows the Seq2Seq framework and includes two
main modules, i.e., the encoder module and the decoder module. The encoder
module can be divided into two parts: local (method-level) encoder
and global (class-level) encoder. The former is responsible for
representing the lexical and syntactic information of each method, and the
latter  encodes the semantic information of each class and the relationship
between classes. The decoder integrates the multi-modality representations
from the above encoders through an attention mechanism and generates the final
summary.

\begin{figure*}[t]
\centering
\includegraphics[width=\textwidth]{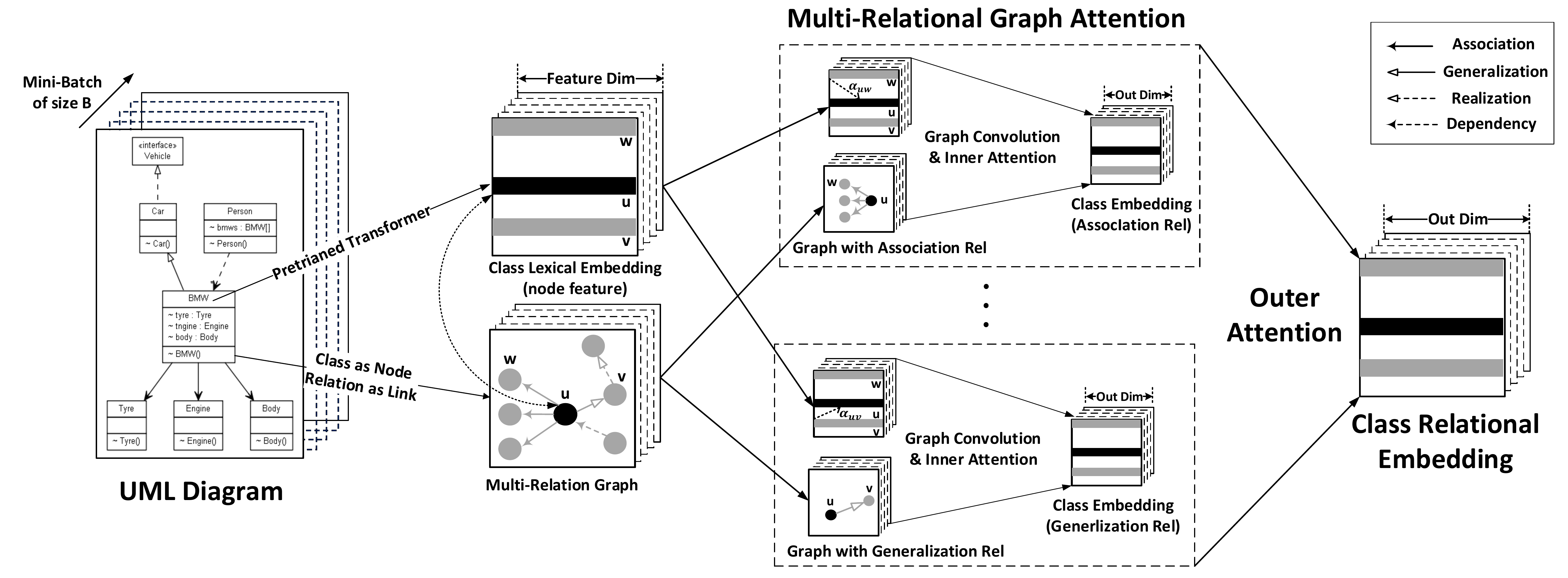}
\vspace{5pt}
\caption{Global context encoder based on our proposed \mrgnn{}} 
\label{fig:MR-GNN}
\end{figure*}

\subsection{Local Context Encoders using GRU}

Contemporary code summarization approaches mainly leverage local contexts
including code tokens and abstract syntax trees (ASTs) to capture the lexical
and syntactic information of a code snippet. For the local context encoders,
we follow the {\astattgru } model~\cite{LeClairJM19}. We preprocess the code
snippets to ASTs and adopt the structure-based traversal (SBT)
format~\cite{HuLXLJ18} to flatten ASTs into sequences. \codeuml{} feeds the code
sequence and SBT sequence into a code encoder and an AST encoder,
respectively. Both encoders adopt GRU~\cite{ChoMGBBSB14}, but the parameters
are not shared:
\begin{equation}\label{eq:gru}
\begin{aligned}
\mathbf{r}_t &= \sigma(\mathbf{W}_{r}\mathbf{x}_t+\mathbf{b}_{r}+\mathbf{W}_{hr}\mathbf{h}_{t-1}+\mathbf{b}_{hr})\\
\mathbf{z}_{t} &= \sigma(\mathbf{W}_{z}\mathbf{x}_t+\mathbf{b}_{z}+\mathbf{W}_{hz}\mathbf{h}_{t-1}+\mathbf{b}_{hz})\\
\mathbf{n}_{t} &= tanh\big(\mathbf{W}_{n}\mathbf{x}_{t}+\mathbf{b}_{n}+\mathbf{r}_{t}\circ(\mathbf{W}_{hn}\mathbf{h}_{t-1} + \mathbf{b}_{hn}) \big)\\
\mathbf{h}_{t} &= (1-\mathbf{z}_t) \circ \mathbf{n}_t + \mathbf{z}_t \circ \mathbf{h}_{t-1}
\end{aligned}
\end{equation}

\noindent where $\mathbf{x}_t$ indicates the embedding of the $t$-th token in
either code token sequence or SBT sequence, $\mathbf{h}_t$ is the hidden state
for the $t$-th token, and $\mathbf{r}_t$, $\mathbf{z}_t$, $\mathbf{n}_t$
indicate the reset, update and new gates, respectively. $\sigma$ denotes the
sigmoid function and ``$\circ$'' is the Hadamard product. $\mathbf{W}_r$,
$\mathbf{W}_{hr}$, $\mathbf{W}_{z}$, $\mathbf{W}_{hz}$, $\mathbf{W}_{n}$ and
$\mathbf{W}_{hn}$ are learnable weight matrices. $\mathbf{b}_{r}$,
$\mathbf{b}_{hr}$, $\mathbf{b}_{z}$, $\mathbf{b}_{hz}$, $\mathbf{b}_{n}$ and
$\mathbf{b}_{hn}$ are bias vectors.

It is worth noting that the global context encoder can benefit to other local context encoders in other methods (e.g., \cite{IyerKCZ16,HuLXLJ18,hu2019deep}), as the local context
encoder is orthogonal to the global context encoder (illustrated in Sec.~\ref{sec:global} and evaluated in Sec.~\ref{sec:RQ4}) in \codeuml.


\subsection{Global Context Encoder based on \mrgnn{}}\label{sec:global}

The global context in \codeuml{} contains the intra-class  (i.e., class name) and the inter-class (i.e., relationship between classes) information. 
For the class-class relationship, we extract the UML class diagram 
that 
the target method resides in and retain four common relationships:
\texttt{Realization}, \texttt{Generalization}, \texttt{Dependency}, and
\texttt{Association}. As illustrated in \Fig \ref{fig:MR-GNN}, the global
context encoder takes two steps to encode global contexts.


Firstly, a class name is preprocessed to a token sequence by the steps
described in \Sec \ref{sec:dataset_pre}. Then \codeuml{} generates class name
embeddings using the Transformer-based sentence embedding
model~\cite{cer2018universal} to obtain the \emph{class semantic embedding}.
The sentence embedding model is trained and optimized for greater-than-word
length text, such as sentences or short paragraphs on a variety of NL data
sources and tasks, and the output is a 512-dimensional vector. The class
semantic embedding is a part of the output of the global context encoder, which
will be passed to the subsequent GNN module as initial features of the nodes
in the UML class diagram.

Secondly, we extract structural relationships from UML and propose 
\mrgnn{} to encode inter-class context.
Specifically, we treat each class as a node and each relationship between two
classes as an edge between corresponding nodes. Since there are four types of
class relationships, we get four graphs accordingly with each graph containing
one of the four relationships. Each class node $v_i$ has one general node
embedding $\mathbf{h}_i^{(g)}$ and four relation-specific embeddings
$\mathbf{h}_i^{(r)}$ with $r$ being one of the four class-class relations. The
class semantic embedding $\mathbf{h}_i^{(l)}$, which is generated by the
sentence embedding model for class node $v_i$, is used as the initial
hidden state for $\mathbf{h}_i^{(g)}$. The UML graphs have certain
characteristics that are rarely considered in general graphs. On one hand,
they have multi-relational edges. On the other hand, the definition of
each type is relatively broad, resulting in a huge  difference in the
semantics of edges with the same relationship. For instance,
\texttt{Association} relationship may exist between \texttt{<Student,
Professor>} and also \texttt{<Student, Course>}, but their  semantics are
obviously different. Some classic GNN models such as GCN~\cite{KipfW17} cannot
handle such scenarios. As a comparison, \mrgnn{} is able to capture
different relation semantics in UML graphs via inner-attention and
outer-attention. The former is mainly used to distinguish the effect of
different neighbors in the same relationship, while the latter reflects the
influence of different relationships.

To be specific, similar to Graph Attention Network
(GAT)~\cite{VelickovicCCRLB18}, inner attention defines an attention layer to
capture the different importance of each neighbor in the same relation $r$:
\begin{equation}\label{eq:gat}
\alpha_{i, j}^{(r)} = \frac{{\rm exp}\Big(\Gamma^{(r)}\big(\mathbf{W}^{(r)}_a \mathbf{h}_i^{(r)} \oplus \mathbf{W}^{(r)}_a \mathbf{h}_j^{(r)})\Big)}{\sum_{k\in \mathcal{N}^{(r)}_i} {\rm exp}\Big(\Gamma^{(r)}\big(\mathbf{W}^{(r)}_a \mathbf{h}_i^{(r)} \oplus \mathbf{W}^{(r)}_a \mathbf{h}_k^{(r)})\Big)},
\end{equation}

\noindent where $\alpha_{i,j}^{(r)}$ is the inner attention coefficient of
node $v_j$ to node $v_i$, $\mathbf{h}^{(r)}_i$ is the hidden representation of
node $v_i$ for relation $r$, $\mathcal{N}_i^{(r)}$ is the first-order
neighbors of node $v_i$ with the relation $r$, ``$\oplus$'' is the
concatenation operation, $\Gamma^{(r)}(\cdot)$ is a single-layer feedforward
neural network with a scalar being the output, and $\mathbf{W}_a^{(r)}$ is a
learnable parameter matrix. Given the inner attention coefficient, the hidden
representation of node $v_i$ with respect to the relation $r$ is:
\begin{equation}
\mathbf{h}^{(r)}_{i} = \sigma\bigg(\sum_{j\in \mathcal{N}_i^{(r)}} \alpha_{i,j}^{(r)}\mathbf{W}_{h}^{(r)}\mathbf{h}_j^{(r)}\bigg),
\end{equation}

\noindent where $\mathbf{W}_{h}^{(r)}$ is the learnable weight matrix and
$\sigma(\cdot)$ is LeakyReLU. 

Outer attention, defined at the relation level, is to aggregate the hidden
representations of different relations for each node. The outer attention
coefficient of relation $r$ to node $v_i$ is defined as:
\begin{equation}
\label{equ:attention2}
\beta^{(r)}_i = \frac{{\rm exp}\big(<\mathbf{W}_{b} \mathbf{h}_i^{(g)}, \mathbf{W}^{(r)}_{c} \mathbf{h}^{(r)}_i>\big)}{\sum_{r\in \mathcal{R}}{\rm exp}\big(<\mathbf{W}_{b} \mathbf{h}_i^{(g)}, \mathbf{W}^{(r)}_{c} \mathbf{h}^{(r)}_i>\big)},
\end{equation}

\noindent where ``$<\cdot \;\, ,\;\cdot>$'' indicates inner product,
$\mathbf{W}_{b}$ and $\mathbf{W}_{c}^{(r)}$ are parameter matrices to ensure
space alignment, and $\mathcal{R}$ is the set of the four relationships in UML
graphs. Based on the outer attention coefficient, we calculate the updated
general node representation as: \begin{equation} \mathbf{h}^{'(g)}_{i} =
\sigma\bigg(\sum_{r\in \mathcal{R}} \beta_{i}^{(r)} \mathbf{W}^{(r)}_{d}
\mathbf{h^{(r)}_i}\bigg), \end{equation}

\noindent where $\mathbf{h}^{'(g)}_i$ is the output representation of node
$v_i$ in the current \mrgnn{} layer and also the input of the next \mrgnn{} layer
when stacking multiple layers, and $\mathbf{W}^{(r)}_{d}$ is a learnable
parameter matrix.  Compared to using concatenation and feedforward network in
\Eq~\ref{eq:gat}, we simply adopt an inner product in
\Eq~\ref{equ:attention2} due to the small number of UML relationship types.
It will help avoid introducing too many parameters and thus causing
over-fitting when calculating the attention coefficient. In practice, we stack
two \mrgnn{} layers to obtain the \emph{class relational embedding}.

\subsection{Attention-based Decoder}

Similar to the code token encoder and AST encoder, we deploy GRU as the
backbone of the decoder. Additionally, we use a two-level attention mechanism
to endow \codeuml{} with the ability to learn from code token sequence, SBT
sequence and UML graph selectively.

Let $\mathbf{H}^{(c)}=[ \mathbf{h}^{(c)}_{1}, \cdots,
\mathbf{h}^{(c)}_{T_c}]$, $\mathbf{H}^{(a)}=[ \mathbf{h}^{(a)}_{1}, \cdots,
\mathbf{h}^{(a)}_{T_a}]$ be the outputs from code token encoder and AST
encoder, where $T_c$ and $T_a$ are the length of code token sequence and SBT
sequence, respectively. Firstly, the hidden state $\mathbf{h}_{t}^{(s)}$ of
the $t$-th summary token is obtained similar to \Eq~\ref{eq:gru}. Then the
sequential attention~\cite{LuongPM15} is used to incorporate the different
coefficients between each token in code token sequence (and SBT sequence) and
the $t$-th summary token, and generate the code context vector
$\mathbf{s}^{(c)}_{t}$ and SBT context vector $\mathbf{s}^{(a)}_{t}$ for the
$t$-th summary token:
\begin{equation}
\small
\begin{aligned}
\mathbf{s}_{t}^{(c)} &= \sum_{i=1}^{T_c} \gamma_{i, t} \cdot \mathbf{h}_i^{(c)}, \quad  \gamma_{i, t} = \frac{{\rm exp}\big(<\mathbf{h}_i^{(c)},  \mathbf{h}_t^{(s)} >\big)}{\sum_{k=1}^{T_{c}}{\rm exp}\big(<\mathbf{h}^{(c)}_k,  \mathbf{h}_t^{(s)}>\big)} \\
\mathbf{s}_{t}^{(a)} &= \sum_{i=1}^{T_a} \delta_{i, t} \cdot \mathbf{h}_i^{(a)}, \quad \delta_{i, t} = \frac{{\rm exp}\big(<\mathbf{h}_i^{(a)},  \mathbf{h}_t^{(s)}>\big)}{\sum_{k=1}^{T_{a}}{\rm exp}\big(<\mathbf{h}^{(a)}_k,  \mathbf{h}_t^{(s)}>\big)}.
\end{aligned}
\end{equation}





We then perform a multi-modal attention to integrate code context, SBT context, intra-class level context $\mathbf{h}^{(l)}$ (i.e., class semantic embedding) and inter-class level context $\mathbf{h}^{(g)}$ (i.e., class relational embedding). The integrated encoding vector $\mathbf{q}_t$ for $t$-th summary token can be obtained by the following formula:
\begin{equation}
\small
\mathbf{q}_{t} = \sigma\big( \lambda^{(c)}_{t} \mathbf{V}_{c} \mathbf{s}_t^{(c)} + \lambda^{(a)}_{t} \mathbf{V}_{a} \mathbf{s}_t^{(a)} + \lambda^{(l)}_{t} \mathbf{V}_{l} \mathbf{h}^{(l)} + \lambda_{t}^{(g)} \mathbf{V}_{g} \mathbf{h}^{(g)}\big),
\end{equation}
where $\mathbf{\lambda}_t = \frac{1}{Z_t}(\lambda^{(c)}_{t}, \lambda^{(a)}_{t}, \lambda^{(l)}_{t}, \lambda_{t}^{(g)})^{\rm T}$ is a normalized attention coefficient vector with $Z_t$ being the normalized factor, and $\mathbf{V}$ denotes a learnable parameter matrix.
To be specific, each entry of vector $\mathbf{\lambda}_t$ is defined as:
\begin{equation}
    \left(
    \begin{array}{c}
        \lambda^{(c)}_{t} \\
        \lambda^{(a)}_{t} \\
        \lambda^{(l)}_{t} \\
        \lambda^{(g)}_{t}
    \end{array}
    \right) = {\rm exp} \left(
    \begin{array}{c}
        <\mathbf{\mathcal{V}}_c \mathbf{s}_t^{(c)}, \mathbf{\mathcal{V}}_s \mathbf{h}_t^{(s)}> \\
        <\mathbf{\mathcal{V}}_a \mathbf{s}_t^{(a)}, \mathbf{\mathcal{V}}_s \mathbf{h}_t^{(s)}> \\
        <\mathbf{\mathcal{V}}_l \mathbf{h}_t^{(l)}, \mathbf{\mathcal{V}}_s \mathbf{h}_t^{(s)}> \\
        <\mathbf{\mathcal{V}}_g \mathbf{h}_t^{(g)}, \mathbf{\mathcal{V}}_s \mathbf{h}_t^{(s)}>
    \end{array}
    \right),
\end{equation}
where $\mathbf{\mathcal{V}}_c, \mathbf{\mathcal{V}}_s, \mathbf{\mathcal{V}}_a, \mathbf{\mathcal{V}}_g$ and $\mathbf{\mathcal{V}}_l$ are parameter matrices to ensure space alignment.

Given the $t$-th summary token hidden state $\mathbf{h}_t^{(s)}$ and the corresponding integrated encoding vector $\mathbf{q}_{t}$, \codeuml{} predicts the probability distribution of the $t$-th summary token as:
\begin{equation}
    \mathbf{y}_t = {\rm softmax} \big(f(\mathbf{h}_t^{(s)} \oplus \mathbf{q}_{t})\big),
\end{equation}
where ``$\oplus$'' is the concatenation operation and $f(\cdot)$ is a single-layer feedforward neural network. The cross entropy loss is utilized to evaluate the gap between the prediction and the ground truth, i.e.
\begin{equation}
    l = \frac{1}{T_s}\sum_{t=1}^{T_s} <\mathbf{\hat{y}}_t, \log \mathbf{y}_t>,
\end{equation}
where $\mathbf{\hat{y}}_t$ is the $t$-th target token by one hot encoding and $T_s$ is the length of the target summary.
Many gradient descent based methods can be used for optimizing \codeuml{} and we adopt the adaptive moment estimation method (i.e., AdamW)~\cite{loshchilov2017decoupled}.
\section{Experimental Setup}\label{sec:exp}


\subsection{Datasets and Preprocessing}
\label{sec:dataset_pre}


In our experiment, we use two Java datasets: CodeSearchNet~\cite{abs-1909-09436} and \codenet (short for \underline{Co}de \underline{Co}mments dataset crawled from inter\underline{Net}) . 
\emph{CodeSearchNet} is a public dataset containing $542,991$ Java methods across the training, validation and test sets. 
In order to build the UML class diagrams for the programs, project-level source code is needed.
Fortunately, the CodeSearchNet dataset provides the \texttt{url} and \texttt{repo} information for each function. 
We download all the projects (publicly available open-source non-fork GitHub repositories) according to the URLs provided in the data. Then, we filter the dataset according to a set of rules:
\begin{enumerate}
\item Code snippets with missing GitHub repositories are removed.
\item Code snippets that cannot be processed by \umlgraph{} to build UML class diagrams are removed. 
\item Summaries shorter than three words are removed.
\item Non-English summaries are removed. 
\end{enumerate} 

After the above preprocessing steps, we get $267,047$ pairs of source code and summaries. \Tab~\ref{tab:num:csn} shows the statistics of the filtered corpus. Following the original partition given in~\cite{abs-1909-09436}, the relative size of train:validation:test is set to be 9:0.4:0.6. We have checked that there is no duplication inside or across the training, validation, and test set. 

\emph{\codenet} is a large-scale dataset that we collected from GitHub public repositories. We collect Java repositories created during 2015 to 2020 and get $1.8M$ raw methods. 
Following \citet{HuLXLJ18} and \citet{GuZZK16}, for each method with a comment, we extract the first sentence in its Javadoc as the comment since it typically describes the functionality of the method according to Javadoc guideline~\cite{javadoc}. 
Following \citet{abs-1909-09436} and \citet{HuLXLJ18}, we filter out methods with name `toString', `hashCode', `equals', `finalize', etc and test methods. We also filter out constructors, generated methods and perform dataset deduplication as in \cite{abs-1909-09436}.
After the data filtering steps, $1.2M$ methods are remained at last. 
The training/validation/test sets splitting of both datasets are by project so that code from the same repository can only exist in one partition.
\Tab~\ref{tab:num:csn} shows the statistics of the filtered corpus of \codenet. We have made the dataset public\footnote{\url{https://figshare.com/articles/dataset/codenet_dataset_pkl/13514191}}.

\begin{table}[t]
  \caption{Statistics of the CodeSearchNet and \codenet dataset}
  \centering
  \label{tab:num:csn}
  \begin{tabular}{cccc}
    \toprule
    & \#Repo &\#UML diagram &\#Method\\ \midrule
    CodeSearchNet & 4,116 & 73,856 & 267,047\\ \midrule
    \codenet{} & 56,506 & 162,543 & 1,228,631 \\
  \bottomrule
\end{tabular}
\end{table}

We use {\umlgraph } to extract UML class diagrams for extracting the class-class relationships. 
Four class-class relationships are extracted: realization, generalization, dependency and association (which are represented by \texttt{IMPLEMENTS}, \texttt{EXTENDS}, \texttt{DEPEND} and \texttt{ASSOC} in UMLGraph, respectively). 
We also apply the following rules to preprocess code and summaries in the dataset: 

\begin{enumerate}

\item We split code and summary tokens into subtokens by applying the
spiral ronin algorithm\footnote{
\url{https://github.com/casics/spiral}} to reduce
data sparsity and then all the tokens are transformed to lower cases.

\item Punctuations in summaries are removed.

\item Numerals and string literals are replaced with the generic tokens
\lstinline{<NUM>} and \lstinline{<STRING>}, respectively.

\item We use srcml\footnote{\url{https://www.srcml.org}} to parse
Java methods into ASTs and then flatten them to SBT sequences as
described in~\cite{HuLXLJ18}.

\end{enumerate}

For class names used in global context, we preprocess a class name by: (1)
splitting by camel case or underscores; (2) removing the \texttt{<T>}-like
tokens in template classes; and (3) removing parent class name, e.g.
\texttt{Parent.Child} will be processed to \texttt{Child}. 

\begin{figure}[t]
\centering
\subfloat{
    \includegraphics[width=0.4\textwidth]{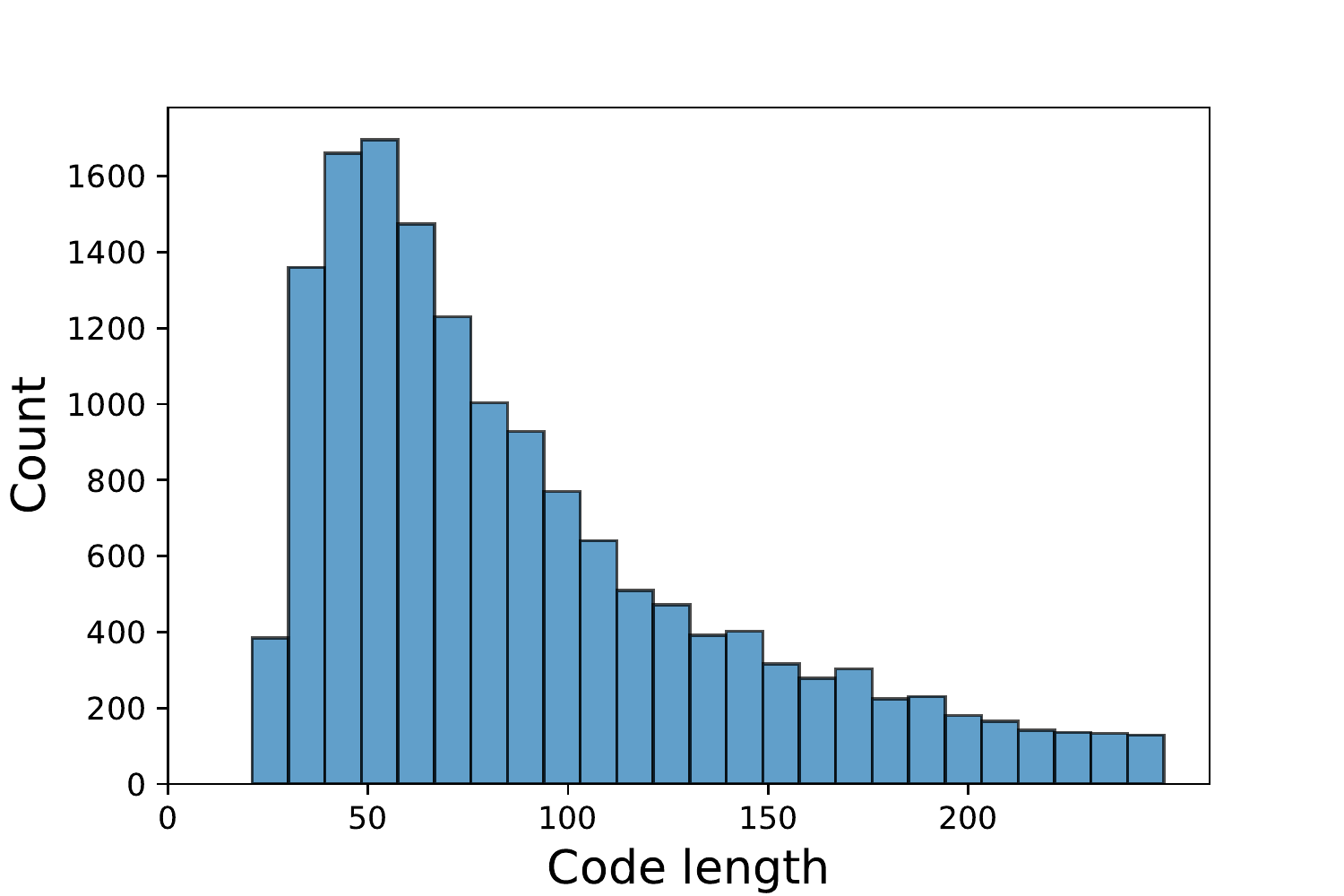}
}
\subfloat{
    \includegraphics[width=0.4\textwidth]{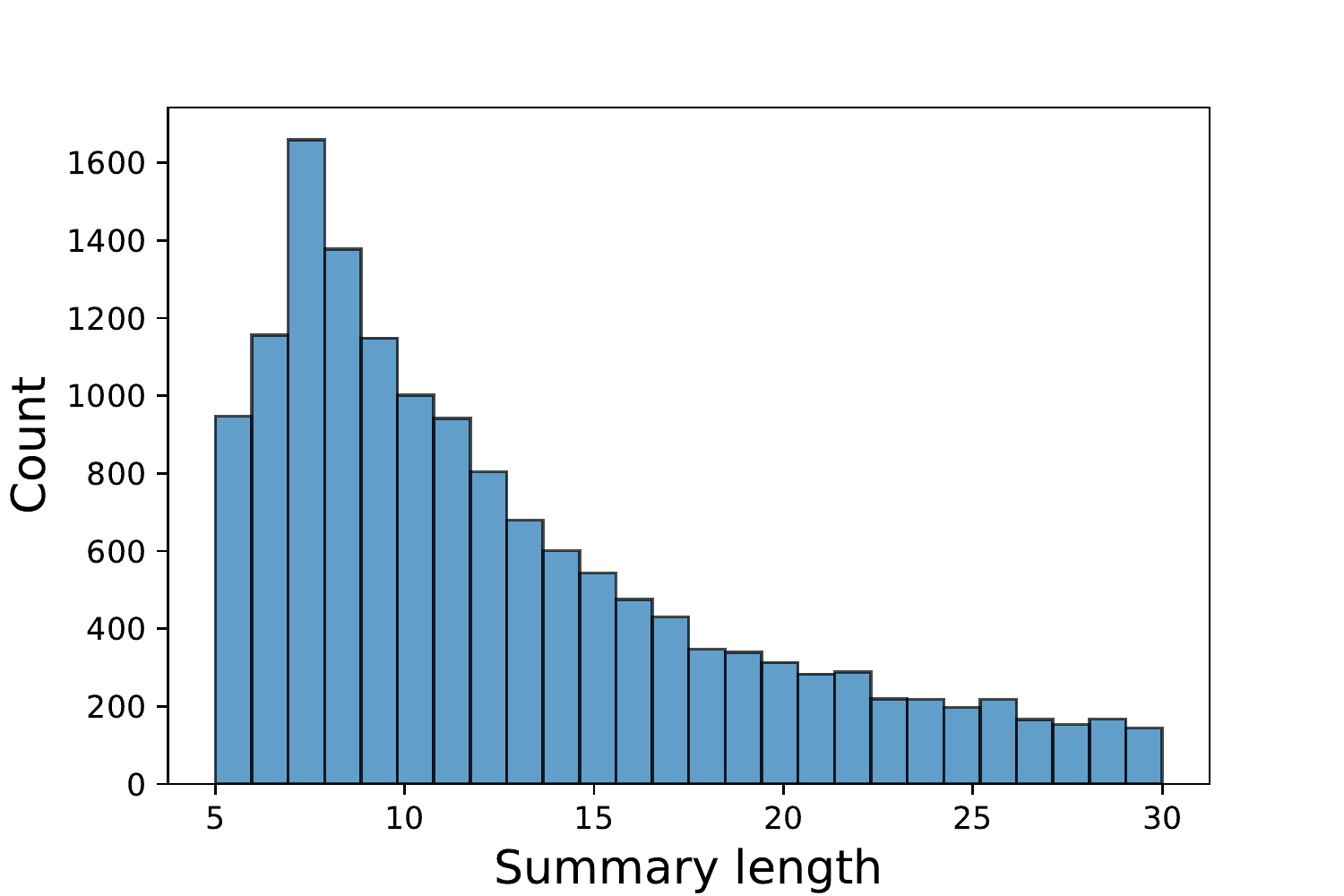}
}
\caption{Length distribution of code and summary}
\label{fig:code_sum_cnt}
\end{figure}

\Fig~\ref{fig:code_sum_cnt} shows the length distribution of code and summaries of CodeSearchNet. 
We can observe that the length of most code snippets ranges from around 30 
to 150 and the length of most summaries ranges from 5 to 15.

\subsection{Experiment Setting}

The vocabulary size is $10,000$ (most frequent words in the training set) for code, summary, and SBT sequences.
For UML diagram, we use a 2-layer \mrgnn{} to encode the class-level
information. The class embedding size and hidden dimensions of \mrgnn{} are $512$ and
$256$, respectively. 
For baseline Ast-attgru~\cite{LeClairJM19}, the embedding size and hidden dimension of GRU are
$128$ and $256$, respectively. We set the mini-batch size to $256$ and a maximum of
$40$ epochs for each approach. We use the optimizer AdamW
\cite{loshchilov2017decoupled} with the learning rate $0.001$ for training. To
prevent over-fitting, we use Dropout with drop probability $0.5$, and Weight
Decay with decay ratio $0.3$. The experiments are conducted on a server with
2 NVIDIA Tesla V100 GPUs.

\subsection{Evaluation Metrics}\label{sec:metrics}

Similar to previous work~\cite{IyerKCZ16,WanZYXY0Y18,zhangretrieval20}, we
evaluate the performance of our proposed model based on four widely-used
metrics including BLEU~\cite{PapineniRWZ02},
METEOR~\cite{BanerjeeL05}, ROUGE-L~\cite{lin-2004-rouge} and
CIDER~\cite{VedantamZP15}. BLEU, METEOR, ROUGE-L and CIDER are prevalent metrics in
machine translation, text summarization and image captioning tasks.

BLEU measures the average n-gram precision between the reference sentences and generated sentences, with brevity penalty for short sentences.  The formula to compute BLEU-1/2/3/4 is:
\begin{equation}
\operatorname{BLEU-N}= BP \cdot \exp \sum_{n=1}^{N} \omega_{n} \log p_{n},
\end{equation}
where  $p_n$ (n-gram precision) is the fraction of n-grams in the generated sentences which are present in the reference sentences, and $\omega_{n}$ is the uniform weight 1/N.
Since the generated summary is very short, high-order n-grams may not overlap. We use the +1 smoothing function~\cite{LinO04}. BP is brevity penalty given as:
\begin{equation}
    BP=\left\{\begin{array}{cl}
                    1 & \text { if } c>r \\
                    e^{(1-r / c)} & \text { if } c \leq r
            \end{array}\right.
\end{equation}
Here, c is the length of the generated summary, and r is the length of  the reference sentence.

Based on longest common subsequence (LCS), ROUGE-L is widely used in text
summarization. Instead of using only recall, it uses F-score which is the
harmonic mean of precision and recall values. Suppose $A$ and $B$ are
generated and reference summaries of lengths c and r respectively, we have:
\begin{equation}
    \left\{\begin{array}{cl}
        P_{ROUGE-L}=\frac{L C S(A, B)}{c}\\
        R_{ROUGE-L}=\frac{L C S(A, B)}{r}\\
     \end{array}\right.
\end{equation}

$F_{ROUGE-L}$, which indicates the value of $\operatorname{ROUGE-L}$, is calculated as the weighted harmonic mean of $P_{ROUGE-L}$ and $R_{ROUGE-L}$:
\begin{equation}
        F_{ROUGE-L}=\frac{\left(1+\beta^{2}\right) P_{ROUGE-L} \cdot R_{ROUGE-L}}{R_{ROUGE-L}+\beta^{2} P_{ROUGE-L}}
\end{equation}
\noindent $\beta$ is set to $1.2$ as in \cite{zhangretrieval20,WanZYXY0Y18}.

METEOR is a recall-oriented metric that measures how well the model captures the 
content from the references in the generated sentences and has a better
correlation with human judgment. Suppose $m$ is the number of mapped unigrams
between the reference and generated sentence with lengths $c$ and $r$ respectively.
Then, precision, recall and F are given as:
\begin{equation}
        P=\frac{m}{c},\,\,
        R=\frac{m}{r},\,\,
        F=\frac{P R}{\alpha P+ (1-\alpha)R}
\end{equation}
The sequence of mapping unigrams between the two sentences is divided into the fewest possible number of ``chunks''. This way, the matching unigrams in each ``chunk'' are adjacent (in two sentences) and the word order is the same. The penalty is then computed as:
\begin{equation}
\text {Pen}=\gamma \cdot \text { frag }^{\beta}
\end{equation}

\noindent where $\text {frag}$ is a fragmentation fraction:
$\text{frag}=ch/m$, where $ch$ is the number of matching chunks and $m$ is
the total number of  matches. The default values of $\alpha, \beta, \gamma$
are 0.9, 3.0 and 0.5 respectively.

CIDER is a consensus-based evaluation metric used in image captioning tasks. The notions of importance and accuracy are inherently captured by computing the TF-IDF weight for each n-gram and using cosine similarity for sentence similarity.
To compute CIDER, we first calculate the TF-IDF weighting $g_k(s_i)$ for each n-gram $\omega_k$ in reference sentence $s_i$. Here $\omega$ is the vocabulary of all n-grams. Then we use the cosine similarity between the generated sentence and the reference sentences to compute $\operatorname{CIDER}_n$ score for n-grams of length $n$. The formula is given as:
\begin{equation}
    \operatorname{CIDER}_{n}\left(c_{i}, s_{i}\right)==\frac{<\boldsymbol{g}^{\boldsymbol{n}}\left(c_{i}\right), \boldsymbol{g}^{\boldsymbol{n}}\left(s_{i}\right)>}{\left\|\boldsymbol{g}^{n}\left(c_{i}\right)\right\|\left\|\boldsymbol{g}^{n}\left(s_{i}\right)\right\|}
\end{equation}
where $\boldsymbol{g}^{\boldsymbol{n}}(s_i)$ is a vector formed by $g_k(s_i)$ corresponding to all the n-grams (n varying from 1 to 4). $c_i$ is the $i^{th}$ generated sentence.
Finally, the scores of various n-grams can be combined to calculate CIDER as follows:
\begin{equation}
\operatorname{CIDER}\left(c_{i}, s_{i}\right)=\sum_{n=1}^{N} w_{n} \operatorname{CIDER}_{n}\left(c_{i}, s_{i}\right)
\end{equation}

Note that we usually report the scores of BLEU, METEOR and ROUGE-L in
percentages since they are in the range of $[0,1]$. As CIDER scores range in
$[0,10]$, we display them in real values.
\section{Evaluation}\label{sec:eval}

In this section, we evaluate \codeuml{} by answering four research questions in Sec. \ref{sec:RQ1} to Sec. \ref{sec:RQ4}. 
We also describe a case study in Sec. \ref{sec:casestudy} and a human evaluation in Sec. \ref{sec:human_eval}. 


\subsection{RQ1: What is the effectiveness of \codeuml{}?}
\label{sec:RQ1}

We evaluate the effectiveness of \codeuml{} by comparing it to the recent work
on code summarization.

\begin{itemize}

\item \textbf{\codenn{}}~\cite{IyerKCZ16} is the first neural approach that
learns to generate summaries of code snippets. It is a classical 
encoder-decoder framework in NMT that encodes
code to context vector with attention mechanism and then generates
summaries in decoder.

\item \textbf{\hybriddrl{}}~\cite{WanZYXY0Y18} is an improved approach with
hybrid code representations (with ASTs) and deep reinforcement learning. It
encodes the sequential and structural content of code by LSTMs and tree-based
LSTMs and uses a hybrid attention layer to get an integrated representation. 

\item \textbf{\hdeepcom{}}~\cite{hu2019deep} is the SBT-based model, which is 
more capable of learning syntactic and structure information of Java methods.

\item \textbf{\astattgru{}} and \textbf{\attgru{}}~\cite{LeClairJM19} are
 encoder-decoder network using GRUs with attention. \astattgru{} is a
multi-encoder neural model that encodes both code and AST.

\item \textbf{\codetoseq{}}~\cite{AlonBLY19} represents a code snippet as a
set of  paths in AST and uses attention to select the relevant paths while
decoding.

\item \textbf{\astnn{}}~\cite{ZhangWZ0WL19} splits large ASTs into sequences
of small statement trees, and encodes the trees to vectors by capturing the
lexical and syntactical knowledge of statements. We applied the model to code
summarization task (original tasks in \astnn{} are source code classification
and code clone detection). 

\item \textbf{\rencos{}}~\cite{zhangretrieval20} enhances the neural model
with the most similar code snippets retrieved from the training set.
Therefore, it leverages both neural and retrieval-based techniques. 

\item \textbf{\neuralcs{}}~\cite{AhmadCRC20} models code with Transformer to capture the long-range dependencies and incorporates the copy mechanism~\cite{SeeLM17} to allow both generating words from vocabulary and copying from the input source code.


\end{itemize}

For baseline comparisons, we re-implement the models of \codenn{} and \hdeepcom{} according to the corresponding papers and use the publicly released code for \hybriddrl{}, \attgru{}, \astattgru{}, \codetoseq{}, \astnn{}, \rencos{} and \neuralcs{}.

\begin{figure*}[t]
    \centering
        \begin{minipage}[t]{0.5\linewidth}
        \centering
        \includegraphics[width=\linewidth]{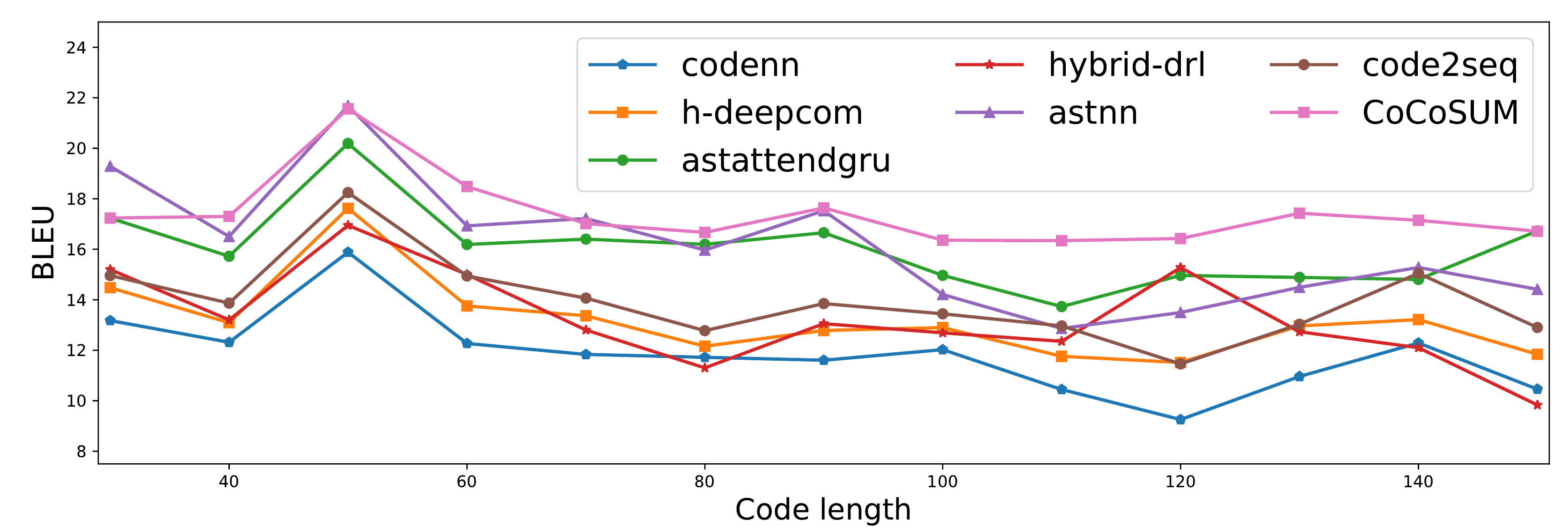}
    \end{minipage}%
    \begin{minipage}[t]{0.5\linewidth}
        \centering
        \includegraphics[width=\linewidth]{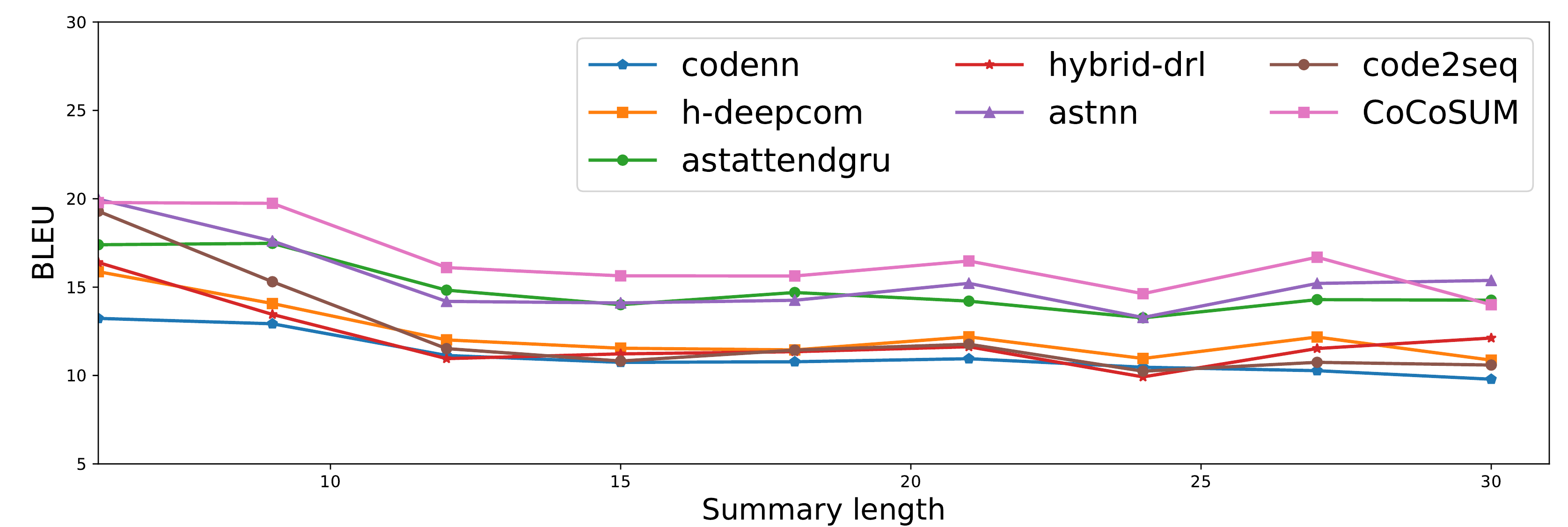}
    \end{minipage}
    \begin{minipage}[t]{0.5\linewidth}
        \centering
        \includegraphics[width=\linewidth]{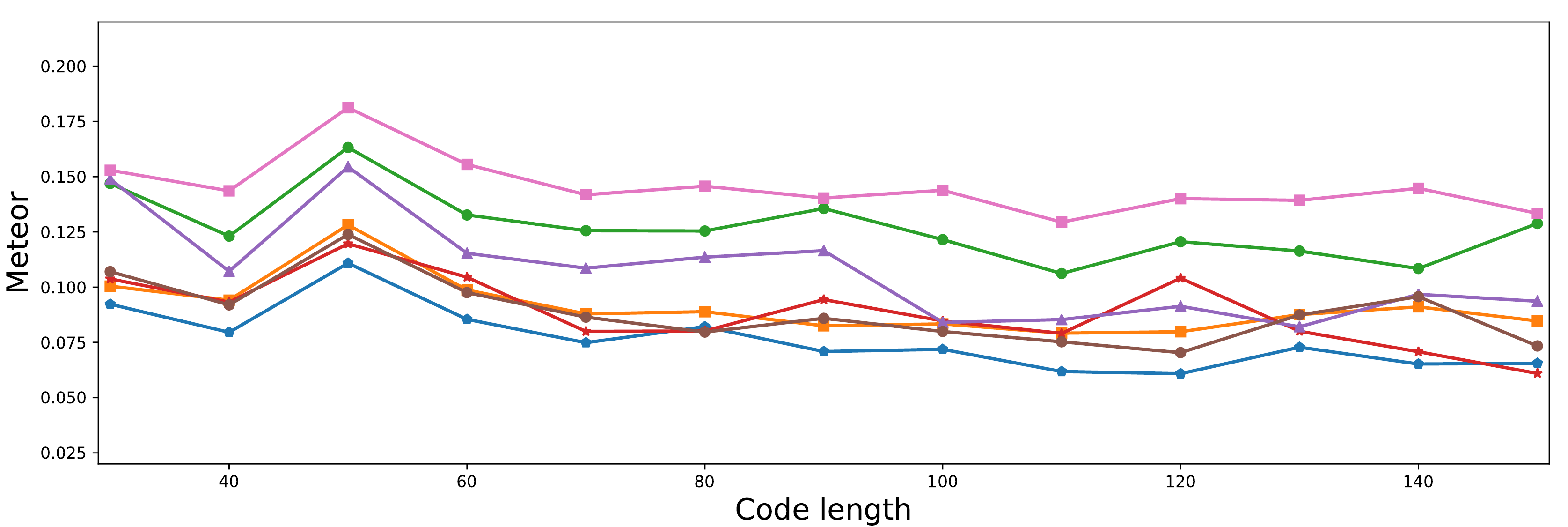}
    \end{minipage}%
    \begin{minipage}[t]{0.5\linewidth}
        \centering
        \includegraphics[width=\linewidth]{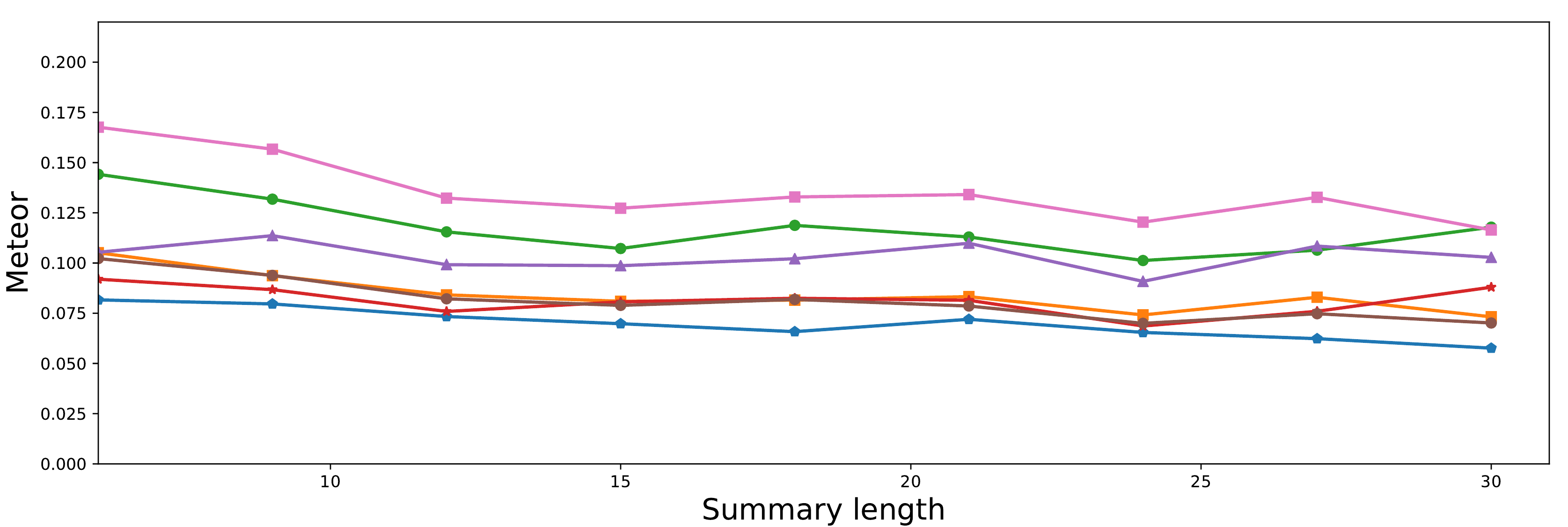}
    \end{minipage}
    
    \begin{minipage}[t]{0.5\linewidth}
        \centering
        \includegraphics[width=\linewidth]{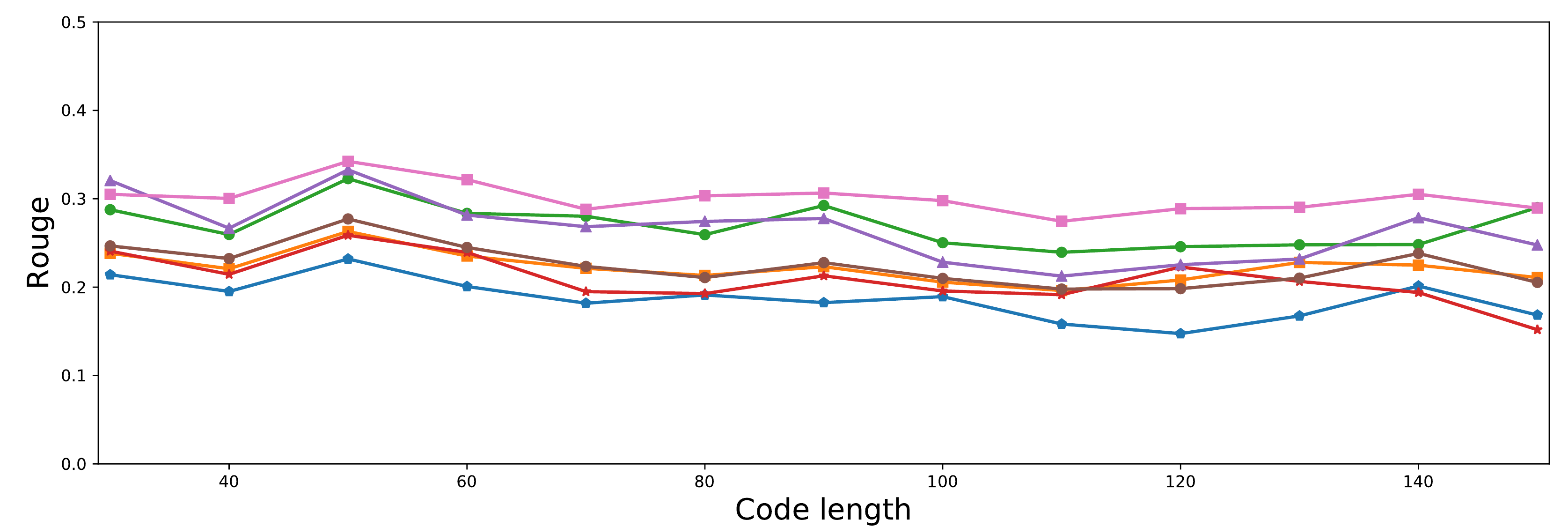}
    \end{minipage}%
    \begin{minipage}[t]{0.5\linewidth}
        \centering
        \includegraphics[width=\linewidth]{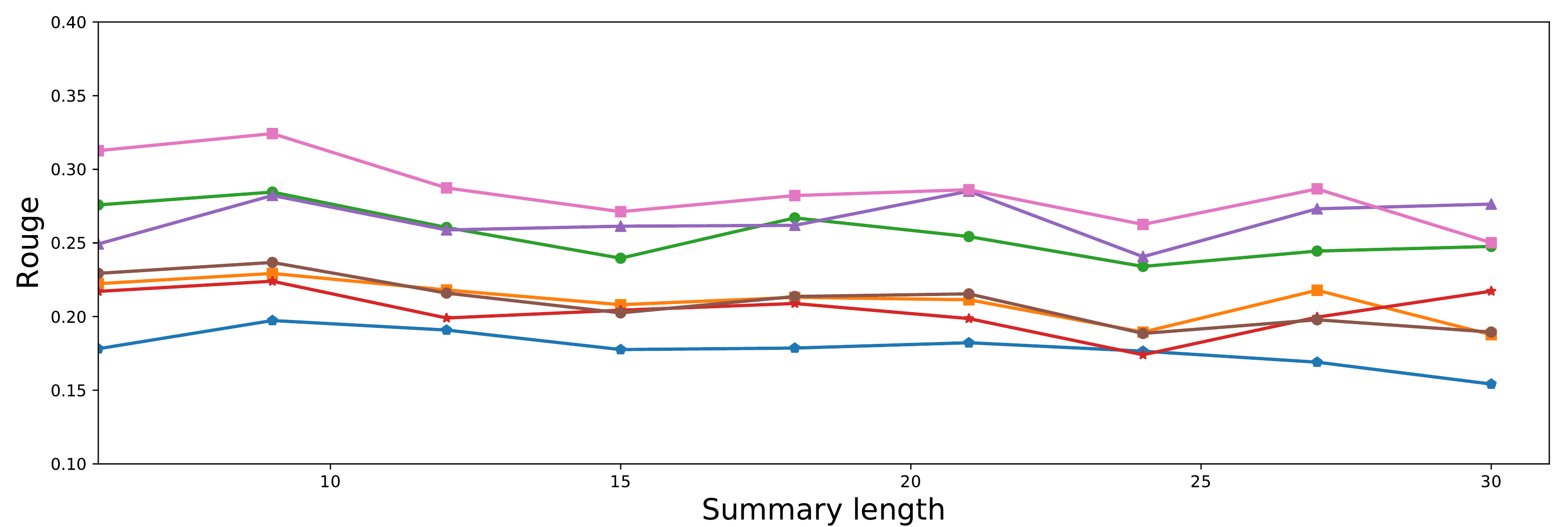}
    \end{minipage}
    \begin{minipage}[t]{0.5\linewidth}
        \centering
        \includegraphics[width=\linewidth]{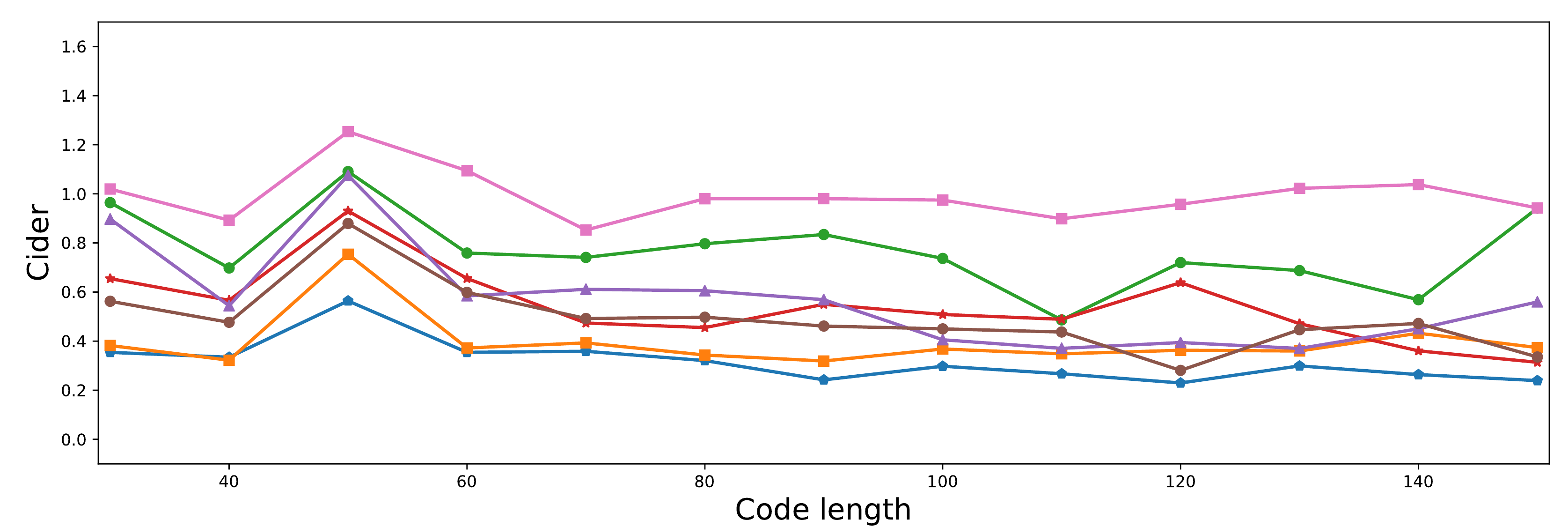}
    \end{minipage}%
    \begin{minipage}[t]{0.5\linewidth}
        \centering
        \includegraphics[width=\linewidth]{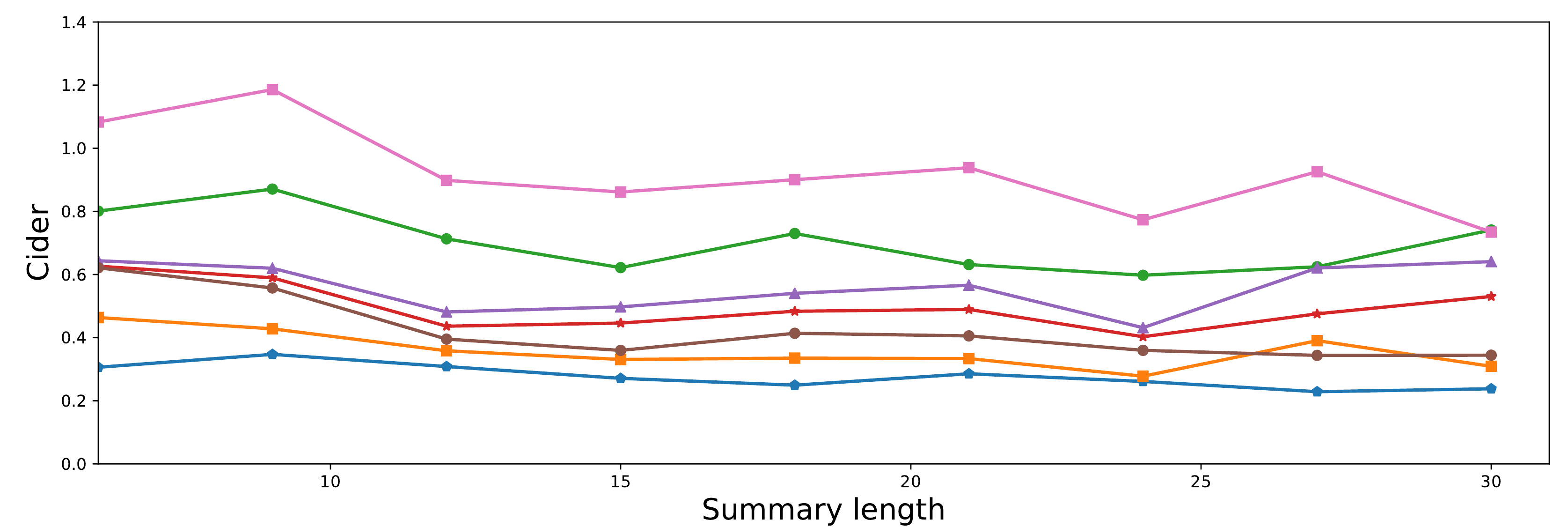}
    \end{minipage}
\caption{Comparison on different code lengths and summary lengths}
\label{fig:diff_code_sum_len}
\end{figure*}

\begin{table}[t]
\caption{Comparison with baselines on CodeSearchNet. The percentages in parentheses indicate the improvement of \codeuml over the best baseline. }
\label{tab:cmp_base}
\centering
\begin{tabular}{ccccc}
\toprule
MODEL&BLEU-4 &METEOR &ROUGE-L &CIDER\\
\midrule
\codenn{}      &12.97  &7.76    &19.54 &0.42 \\
\hybriddrl{} & 14.14 &  8.47 &21.09 & 0.64 \\
\hdeepcom{}   &  14.48    &9.20  &22.95 &  0.51\\
\attgru{}     & 17.05    &12.18   &27.20 &  0.85\\
\astattgru{}   & 17.19    &12.43 &27.52 & 0.90 \\
\codetoseq{} &15.02 &8.98 &23.38 & 0.61 \\
\astnn &17.62 &10.72  &27.76 &0.70 \\
\rencos &15.79 &12.92 &23.35 &0.82 \\
\neuralcs &14.54 &10.41&27.97& 0.82 \\
\midrule
\multirow{2}{*}{\textbf{\codeuml{}}} & \textbf{19.04} &\textbf{14.18} &\textbf{30.77} &\textbf{1.15}\\
            & \textbf{(8.06\%)} & \textbf{(9.75\%)} & \textbf{(10.01\%)} & \textbf{(27.78\%)} \\
\bottomrule
\end{tabular}
\end{table}



\begin{table}[t]
\caption{Comparison with baselines on \codenet. The percentages in parentheses indicate the improvement of \codeuml over the best baseline.} 
\label{tab:cmp_base_codenetfull}
\centering
\begin{tabular}{ccccc} \toprule
MODEL       &BLEU-4 &METEOR &ROUGE-L &CIDER \\ \midrule
\codenn     &13.02  &8.90   &20.72   &0.46 \\
\hdeepcom   &16.29  &11.86  &26.40   &0.74 \\
\attgru     &18.55  &13.80  &29.28   &1.11 \\
\astattgru  &18.91  &13.88  &29.48   &1.14  \\
\codetoseq  &14.46  &9.31   &23.15   &0.64 \\ 
\astnn      &16.79  &9.42   &25.43   &0.52 \\
\rencos     &19.18  & 14.85 &27.56   &1.21 \\
\neuralcs   &16.51  &8.60   &28.80   &1.17  \\ \midrule
\multirow{2}{*}{\textbf{\codeuml}} & \textbf{20.15} & \textbf{14.94} &\textbf{31.24} &\textbf{1.30}\\
            &\textbf{(5.06\%)} &\textbf{(0.61\%)} &\textbf{(5.97\%)} &\textbf{(7.44\%)} \\
\bottomrule
\end{tabular}
\end{table}



Table~\ref{tab:cmp_base} shows the performance of different models on CodeSearchNet dataset in terms of four metrics. We compute BLEU scores in the same way as
\codenn{}~\cite{IyerKCZ16} and the METEOR, ROUGE-L, CIDER scores as
\hybriddrl{}~\cite{WanZYXY0Y18}.
From the table, we can see that \codeuml{} achieves the best scores as marked in
bold and improves the best baseline by 8.06\% to 27.78\% in four metrics. 
Among all the models, \codenn{} is the weakest because it only uses raw
code token sequences and fails to capture the semantic and syntax information.
\hybriddrl{} is better than \codenn{} because it uses AST information and
applies reinforcement learning. However, it uses Tree-based LSTM to model the
whole tree, leading to gradient vanishing and slow training. It costs around
20 days to train and evaluate this baseline on CodeSearchNet dataset.
Therefore, we will omit the comparison to \hybriddrl{} in other experiments.
\hdeepcom{} applies SBT to capture semantic and structural information but the
encoded information cannot be fully utilized in the decoding phase.
\astattgru{} combines several types of information including code and SBT to
improve  summary generation results. 
\astnn{} uses a novel way to split ASTs to better utilize syntactical
information. 
Therefore, \astattgru and \astnn perform relatively well among all baselines.

To verify the generality of applying \codeuml{} on other datasets, we perform the experiments on our newly collected dataset \codenet{}. As shown in Table~\ref{tab:cmp_base_codenetfull}, \codeuml{} also consistently outperforms all the baselines. 
One can observe that the performance of \rencos increases a lot on \codenet{} dataset compared to CodeSearchNet. This is because \rencos{} is a retrieval-based model, as the dataset size increases, it is more likely to retrieve a similar code snippet from the dataset. This observation inspires us to explore adding retrieval module in \codeuml{} in the future work.
In summary, the comparison results show that the proposed model \codeuml{}
employing global contexts is effective with respect to the four quantitative
metrics.

\subsection{RQ2: How much do different components/techniques contribute?}

\subsubsection{Global context ablation study}

\begin{table}[t]
\caption{Global contexts ablation study}
\label{tab:global_contexts_ablation}
\centering
\begin{tabular}{ccccc} \toprule
MODEL       &BLEU-4 &METEOR &ROUGE-L &CIDER \\ \midrule
\astattgru  & 17.19 &12.43  &27.52   &0.90 \\ \midrule
\bertatt    &17.28  &12.32  &27.49   &0.89 \\
\codeumlo   & 18.82 &13.83  &30.29   &1.12 \\
\textbf{\codeuml} & \textbf{19.04} &\textbf{ 14.18 } &\textbf{30.77} &\textbf{1.15}\\ \bottomrule
\end{tabular}
\end{table}

We implement global contexts by adding the class name and UML channels on top
of \astattgru{}. We do channel ablation study by comparing \bertatt{}
(employing intra-class context, \codeuml{} without UML channel) and
\codeumlo{} (employing inter-class context, \codeuml{} without class name
channel) with \codeuml{} (employing both).
Table~\ref{tab:global_contexts_ablation}  shows that both class name and UML
channels improve over \astattgru{}. Since the functionality of a method is
related to its corresponding class, \bertatt{} uses the information of class
names to help generate summaries. 
\codeumlo{} uses the inter-class context information by the UML channel to
capture the $\left<class, class\right>$ relationship and generate better
summaries of source code.  We first obtain the embedding of each node based on
the class information and UML graph structure. Then we encode classes based on
\mrgnn{} and eventually get the representation of all nodes via the attention
mechanism.  \codeuml{} as the combination of all channels, further improves
\bertatt{} and \codeumlo{}, with the UML channel contributing more to the results.


\subsubsection{Multi-relational v.s. Single-relational GNN}

We use multi-relational GNN model in \codeuml{}, which treats different edges
(class relationships) differently. This is because we assume that in UML class diagrams,
different neighbor classes of a target class (i.e., the class enclosing the
target method) via different relationships may have different impact on
summary generation. 

To verify the effectiveness of such design, we conduct an ablation study with a single-relational GNN (\codeumlh{}) that has the same architecture as \codeuml{} except for treating different class relationships as the same type.

The results in Table~\ref{tab:heterogeneous_ablation} show that multi-relational model
outperforms the single-relational one, which confirms our assumption.

\begin{table}[t]
\caption{Multi-relational v.s. single-relational graph model}
\label{tab:heterogeneous_ablation}
\centering
\begin{tabular}{ccccc} \toprule
MODEL       &BLEU-4 &METEOR &ROUGE-L &CIDER \\ \midrule
\codeumlh   &18.88  &14.11  &30.31   &1.13  \\
\textbf{\codeuml} & \textbf{19.04} &\textbf{ 14.18 } &\textbf{30.77} &\textbf{1.15}\\ \bottomrule
\end{tabular}
\end{table}



\subsection{RQ3: What is the stability of \codeuml{}?}

\begin{table*}[t]
\caption{Case studies}
\label{tab:case_1}
\centering
\begin{tabular}{l|l}
\toprule
Case 1 & \begin{lstlisting}
public FieldVisitor visitField(
        final int access,
        final String name,
        final String descriptor,
        final String signature,
        final Object value) {
    if (cv != null) {
        return cv.visitField(access, name, descriptor, signature, value);
    }
    return null;
}

 

Enclosing Class Name: §\colorbox{yellow}{ClassVisitor}§
\end{lstlisting} \\ \hline
Ground Truth & \emph{visits a field of the \textcolor{blue}{class}}\\ \hline
\textbf{\codeuml{}} & \textbf{visit a field in the \textcolor{blue}{class}} \\ \hline
\bertatt{} & visit a field in the given \textcolor{blue}{class} \\ \hline
\astattgru{} & this method is called when a constant is a member of\\ \hline
\attgru{} & this method is called when a constant is a member of \\ \hline
\hdeepcom{} & visit a visitor for a the\\ \hline
\hybriddrl{} & adds the visitor to the visitor \\ \hline
\codenn{} & create the visitor to the the the\\ \hline
\neuralcs{} &visit a field visitor \\ \hline
Case 2 & \begin{lstlisting}
public PropertyStatus getProperty(QualifiedName propertyName) throws DAVException {
    Collection names = new HashSet();
    names.add(propertyName);
    URLTable result = getProperties(names, IContext.DEPTH_ZERO);
    URL url = null;
    try {
        url = new URL(locator.getResourceURL());
    } catch (MalformedURLException e) {
        throw new SystemException(e);
    }
    Hashtable propTable = (Hashtable) result.get(url);
    if (propTable == null)
        throw new DAVException(Policy.bind("exception.lookup", url.toExternalForm()));
    return (PropertyStatus) propTable.get(propertyName);
}

 

Enclosing Class Name: §\colorbox{yellow}{AbstractResourceHandle}§
UML (partial): ['AbstractResourceHandle','§\colorbox{yellow}{PropertyStatus}§',{'relationtype': '§\colorbox{green}{DEPEND}§'}] 
\end{lstlisting} \\ \hline
Ground Truth & \emph{return the \textcolor{blue}{property status} for the property with the given name}\\ \hline
\textbf{\codeuml{}} & \textbf{return a \textcolor{blue}{property status} object for the given property name} \\ \hline
\bertatt{} & get a property from the given property \\ \hline
\astattgru{} & get the property value of the property\\ \hline
\attgru{} & get the property value of the property \\ \hline
\hdeepcom{} & get the property property of the the the\\ \hline
\hybriddrl{} & return the property value\\ \hline
\codenn{} & get the property of the the the the\\ \hline
\neuralcs{}&replies the property status for the given property name \\ \bottomrule
\end{tabular}
\end{table*}



We analyze the stability of \codeuml{} on code and comments of different
lengths. \Fig~\ref{fig:code_sum_cnt} displays the length distribution of code
and summary on testing set. The lengths of most code are less than 150 and the
lengths of most summaries are less than 30, so we conduct the experiment to
evaluate the performance for code lengths varying from 1 to 150 and summary
lengths varying from 1 to 30.

\Fig~\ref{fig:diff_code_sum_len} presents the performance of \codeuml{} and
other models for code and summaries of different lengths. 
The first column of \Fig~\ref{fig:diff_code_sum_len} displays the results of all models for varying code lengths on the four metrics, and the second column displays the results of all models for varying summary lengths on the four metrics. We have the following observations:
\begin{itemize}
    \item \codeuml{} performs the best on four metrics with varying code lengths.
    \item \codeuml{} outperforms others with varying summary lengths.
    \item The performance of baselines decreases 
    when code length increases, while our model performs more stable 
    on all metrics even for longer code.

\end{itemize}
We conclude that \codeuml{} achieves the best performance on code and comments of different lengths. The performce of \codeuml{} with differnt code lengths is relatively stable. The performance of \codeuml{} slightly decreases when summary length is increased.

\subsection{RQ4: What is the generality of \codeuml{}?}
\label{sec:RQ4}

We analyze the generality of \codeuml{} by applying global contexts to some
representative models for code summarization, namely \codenn{}, \hdeepcom{},
\transformer{}, and \astattgru{}. We implement it by adding the class names
and UML information to these models just as we did for \codeuml{}. 
Table~\ref{tab:uml_generality} shows the results for all models are improved
accordingly, by the range of $3.2\%$ to $10.8\%$. This verifies the generality
of global contexts. 

Moreover, the increasing ratios are in the same magnitude, meaning that the
improvement caused by employing global contexts is relatively stable. 

The detailed implementation of each compared model is also available online
\footnote{Implementation of compared models is included in the \texttt{model} folder in the supplementary materials}.

\fix{Transformer remove code channel.}

\begin{table}[t]
\caption{A generality study on global contexts}
\vspace{5pt}
\centering
\label{tab:uml_generality}
\begin{tabular}{ccc} \toprule
MODEL       &without global contexts    &with global contexts \\ \midrule
\codenn     &13.37                      &\textbf{14.72 ($\uparrow$ 10.1\%)}\\
\hdeepcom   &15.16                      &\textbf{15.65 ($\uparrow$ 3.2\%)}  \\
\transformer&14.62                      &\textbf{15.60 ($\uparrow$ 6.7\%)} \\
\astattgru  &17.19                      &\textbf{19.04 ($\uparrow$ 10.8\%)} \\ \bottomrule
\end{tabular}
\end{table}

\subsection{Case study}
\label{sec:casestudy}

To conduct a qualitative analysis of our approach, we compare the generated summaries
from different models and present two case studies as shown in Table~\ref{tab:case_1}.
We have the following
observations from the results that demonstrate the superiority of our approach:

\begin{itemize}

\item In general, \codeuml{} and \bertatt{} can generate summaries better than
other baselines and the summaries of our models are more similar to the ground
truth. 

\item Compared to the baselines, \codeuml{} and \bertatt{} can generate
summary tokens that accurately describe the purpose of the method but cannot
be found within the scope of the method. For example, one important summary
token should be generated in case 1 is ``class'' but it  cannot be generated
only based on the local context (source code itself). \codeuml{} and
\bertatt{} also consider the global context (e.g. the class name
``ClassVisitor''). Therefore, they are able to produce better summarization.


\item Case 2 shows that \codeuml{} generates the best summary expressing the
full meaning of the ground truth. Specifically, the ground truth summary
contains a phrase ``property status'' that cannot be generated by other
models. The reason is that the enclosing class of the target method is
\lstinline{AbstractResourceHandle}, which has a \lstinline{DEPEND} relation
with class \lstinline{PropertyStatus}. This information is learned by our
\mrgnn{} module which analyzes the corresponding UML class diagram. One may
notice that although the token `PropertyStatus' also appears in the method
local context, it is hard for other models to capture this information without
the help of global context. The result emphasizes the importance of
class-class relationships. 


\end{itemize}

\subsection{Human evaluation}\label{sec:human_eval}

\begin{figure*}[t]
\vspace{5pt}
\centering
\includegraphics[width=0.9\linewidth, frame]{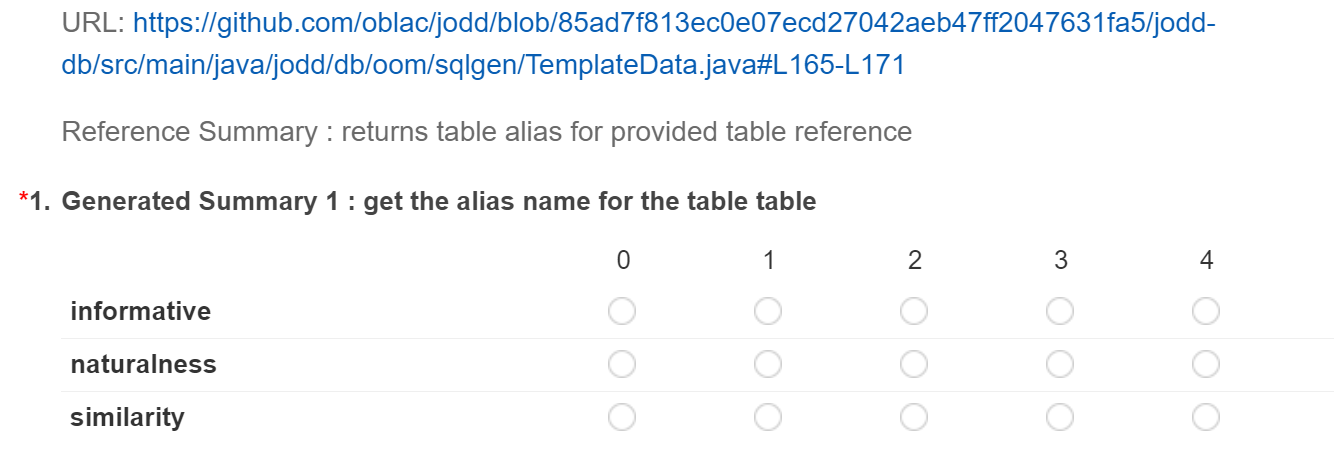}
\caption{A survey question example}
\label{fig:survey}
\end{figure*}
\vspace{5pt}



\begin{table}[t]
\caption{Results of human evaluation (standard deviation in parentheses).} 
\vspace{3pt}
\label{tab:human_evaluation}
\centering
\begin{tabular}{cccc}
\toprule
Model       &Informativeness        &Naturalness            &Similarity \\ \midrule
\codeuml    &\textbf{2.44} (1.25)   &\textbf{3.28} (1.06)   &\textbf{2.16} (1.22) \\  
\astattgru  &2.10 (1.17)            &2.99 (1.04)            &1.94 (1.07)) \\
\astnn      &1.76 (1.05)            &2.00 (1.31)            &1.44 (1.06) \\
\rencos     &1.76 (1.07)            &2.52 (1.19)            &1.58 (1.02) \\  \bottomrule
\end{tabular}
\end{table}
\vspace{5pt}

\begin{table}[t]
\caption{Statistics significance p-value of \codeuml over other methods in human evaluation.} 
\vspace{3pt}
\label{tab:human_evaluation_pvalue}
\centering
\begin{tabular}{cccc}
\toprule
Model       &Informativeness        &Naturalness            &Similarity \\ \midrule
\astattgru  &$1.83e^{-5}$           &$5.20e^{-3}$           &$7.51e^{-3}$ \\
\astnn      &$2.71e^{-13}$          &$3.26e^{-19}$          &$2.30e^{-12}$ \\
\rencos     &$1.72e^{-14}$          &$1.60e^{-10}$          &$5.10e^{-12}$ \\  \bottomrule
\end{tabular}
\end{table}


We also conduct a human evaluation to investigate our approach's effectiveness. We follow the previous work~\cite{IyerKCZ16,LiuXHLXW18,Liu0T0L19,hu2019deep,WeiLLXJ20} to design the human evaluation 
from three aspects: \textbf{similarity} of the generated summary and the ground truth summary, \textbf{naturalness} (grammaticality and fluency of the generated summary), and \textbf{informativeness} (the amount of content carried over from the \emph{input code} to the generated summary, ignoring fluency).
We invite $8$ evaluators to assess the quality of summaries generated by \codeuml{} and three baselines \astattgru{}, \astnn{}, and \rencos{}. The evaluators are software developers or researchers in Company M

with more than 5 years of software development experience and fluent English ability. According to Table~\ref{tab:cmp_base}, the baselines we choose are the top ones among all compared methods (we did not choose \attgru{} since it is a weaker version of \astattgru{}). 

We randomly select 50 source code examples and the corresponding generated 
results of the four models from the test set. As illustrated by \Fig~\ref{fig:survey}, for each
example, we show evaluators its source code link, the ground truth summary, and the generated summary by one of the four models. The evaluators have no idea about which summary is generated by which model. The evaluators are asked to assign scores from 0 to 4 (the higher the better) 
to measure the informativeness, naturalness, and similarity of the generated summary. 
We assign 4 evaluators to each question. 

The evaluation results are shown in \Tab~\ref{tab:human_evaluation}
. Our approach \codeuml outperforms \astattgru, \astnn\footnote{Note that the original \astnn{} model is not designed for the code summarization task. Therefore, the low score does not directly translate to inferior model.} and \rencos in all three aspects, with average scores 2.44, 3.28, and 2.16 for informativeness, naturalness, and similarity, respectively. 
Consistent with \citet{WeiLLXJ20}'s finding, we find that the scores of informativeness of all methods are higher than those of similarity,  indicating that the generated summaries are more relevant to the input code. 


We also conduct Wilcoxon signed-rank tests~\cite{wilcoxon1970critical} for the human evaluation. The results are shown in \Tab~\ref{tab:human_evaluation_pvalue}. Comparing \codeuml with \astattgru{}, \astnn{}, and \rencos{}, all p-values of Wilcoxon signed-rank tests at $95\%$ confidence level are smaller than 0.01,  which mean that the improvements achieved by our approach are statistically significant. In summary, the results of human evaluation confirm the effectiveness of \codeuml.


\section{Threats to Validity}
\label{sec:threats}
We have identified four main threats to validity. 

\begin{itemize}
\item We evaluate and compare our work only on Java datasets. Although in principle, the model is not specifically designed for Java, more evaluations are needed when generalizing our work to other languages.

\item In the design of a neural network model, there are many orthogonal aspects such
as RNN unit type (some papers use GRU, some use LSTM, etc), different token
embedding, teacher forcing, etc. When
showing the generality of \codeuml{}, we have done the experiments in a
controlled way by only adding/removing global contexts. 
An important future work is to explore variant designs of \codeuml such as different token splitting ways, different RNN choice, etc, which has potential to bring further performance improvement.

\item The summaries in CodeSearchNet and \codenet{} datasets are collected by extracting the
first sentences of Javadoc. Although this is the common practice to place a
method's summary at the first sentence according to the Javadoc
guidance\footnote{\href{http://www.oracle.com/technetwork/articles/java/index-137868.html}{http://www.oracle.com/technetwork/articles/java/index-137868.html}} and other works~\cite{GuZZK16,hu2019deep}, there might
still exist some mismatched summaries in the dataset. A higher-quality dataset 
could be used in the future. 


\item In human evaluation, interrater reliability could be a threat to validity: bias may exist in the scores assigned to the same sample by different evaluators. To mitigate this threat, we explain the meaning of each score carefully to the evaluators before actual scoring. Each generated summary is evaluated by 4 human evaluators, and we use the average score of the four evaluators as the final score. As shown in \Tab~\ref{tab:human_evaluation}, the standard deviations of all methods are very small (ranging between 1.02 to 1.31), meaning that their scores are concentrated and close to each other. 

\end{itemize}

\section{Related Work}\label{sec:rw}

\fix{Other related work:
Fernandes et al.~\cite{fernandes2018structured} extended sequence encoders with a graph component modeled by GNN that can reason about long-distance relationships in code.
}

\subsection{Source code representation}

Previous work proposed various representations of source code for follow-up analysis~\cite{AllamanisBBS15,IyerKCZ16,GuZZK16,MouLZWJ16,PiechHNPSG15,ParisottoMSLZK16,DamTP16,LingBGHKWS16,AllamanisTGW15,ZhangWZ0WL19,SuiCZ020}. 
For example, \citet{AllamanisBBS15,SajnaniSSRL16} and \citet{IyerKCZ16}

considered source code as plain text and use
traditional token-based methods to capture lexical information. 
\citet{GuZZK16} used the Seq2Seq model originally introduced for statistical machine translation to learn the intermediate vector representations of queries in natural language and then predict relevant API sequences. 
\citet{MouLZWJ16} proposed a novel Tree-Based Convolutional Neural Network (TBCNN). In TBCNN, program vector representations are learned by the coding criterion; structural features are detected by the convolutional layer; and TBCNN can handle trees with varying children sizes with the continuous binary tree and dynamic pooling. 


\citet{DamTP16} built a language model for modeling software code using LSTMs. 
\citet{LingBGHKWS16} and \citet{AllamanisTGW15} combined the code-context representation with representations of other modalities (e.g., natural language) to synthesize code. 
\citet{AllamanisPS16,HuLXLJ18} apply neural source code representation based on abstract syntax trees and obtain better performance in some programming tasks such as code summarization.

~\citet{AlonZLY19,AlonBLY19} represented a code snippet as a set of compositional
paths in ASTs. 
\citet{ZhangWZ0WL19} proposed an AST-based Neural Network (ASTNN) that splits each large AST into a sequence of small statement trees, and encodes the statement trees to vectors by capturing the lexical and syntactical knowledge of statements and apply the representation to tasks such as source code classification and code clone detection. 
\citet{li2019improving} utilized Program Dependence Graph and Data Flow Graph  as global context in code representation for bug detection. \codeuml{} differs from it in the contexts that are being utilized and in the target task. 
\citet{SuiCZ020} proposed a code embedding method Flow2Vec that embeds control-flows and data-flows of a program in a low-dimensional vector space. The value-flow embedding is formulated as matrix multiplication to preserve context-sensitive transitivity through CFL reachability by filtering out infeasible value-flow paths. Experimental results show that this code representation can improve the performance in code classification and method name prediction task (which was termed as `code summarization task' in their paper). However, their method requires the input program to be compilable
(to LLVM in their experiments) therefore restricts the applicable scenarios, where our method \codeuml has no requirement on compilation.

\subsection{Source code summarization}

Popular source code summarization approaches include rule-based,
IR-based, and learning-based approaches. 

\subsubsection{Rule-based and IR-based approaches}

In the early stage of automatic source code summarization, rule-based and IR-based approaches are widely used~\cite{SridharaHMPV10,HaiducAM10ICSE,HaiducAMM10WCRE, EddyRKC13,RodegheroMMBD14,mcburney2014automatic}. 
For example, \citet{SridharaHMPV10} designed heuristics to choose statements from Java methods and use the Software Word Usage Model (SWUM) to identify keywords from those statements and create summaries though manually-defined templates. The summaries are generated according to the function/variable names via these templates. 
\citet{mcburney2014automatic} 
considered method-level context by choosing important methods that are invoked by the target method via PageRank, and then used SWUM to exact program keywords from the source code and finally generated summaries with a novel natural language generation system. This work uses method-level context (call graph), while we consider different context (class-class relationship).

The basic ideas of IR approaches are retrieving terms from source code for generating term-based summaries or retrieving similar source code and using its summary as the target summary~\cite{HaiducAM10ICSE,HaiducAMM10WCRE,EddyRKC13,RodegheroMMBD14}.
\citet{HaiducAM10ICSE,HaiducAMM10WCRE} used TF-IDF to rank the terms in a method body based on the importance those terms are to that method. TF-IDF gives higher scores to keywords which are common within a particular method body, but rare throughout the rest of the source code. They treat each function of source code as a document and index on such a corpus by LSI or VSM, then the most similar terms based on cosine distances between documents are selected as the summary. 
\citet{EddyRKC13} used topic modeling to improve the work.
\citet{RodegheroMMBD14} use eye-tracking and modify the weights of VSM for better code summarization. 
Code clone detection techniques were also used to retrieve similar code snippets
from a large corpus and reuse their summaries as the targets~\cite{WongYT13,WongLT15}. 

\subsubsection{Learning-based approaches}
\citet{AllamanisBBS15} created the neural logbilinear context model for suggesting method and class names by embedding them in a high dimensional continuous space. 
\citet{AllamanisPS16} also suggested a convolutional model for summary generation, which uses attention over a sliding window of tokens. They summarized code snippets into extreme, descriptive function names. 

The NMT-based models are also widely used for code summarization~\cite{IyerKCZ16,haije2016automatic,HuLXLJ18,HuLXLLJ18,WanZYXY0Y18,hu2019deep,LeClairJM19}. 
\citet{IyerKCZ16} designed a token-based neural model using LSTMs and attention for translation between source code snippets and natural language descriptions.
\citet{haije2016automatic} modeled the code summarization problem as a machine translation task, and some translation models such as Seq2Seq~\cite{SutskeverVL14} and attentional Seq2Seq were employed.
\citet{HuLXLJ18} structurally flattened an AST and then passed it on to a standard Seq2Seq model. Hu et al.~\cite{HuLXLLJ18} leveraged API sequences knowledge to improve summary generation with the learned API sequence knowledge. 
\citet{LeClairJM19} designed a multi-input neural network using GRUs and attention to generate summaries of Java methods, which allows the model to learn code structure independent of the text in code.
\citet{WanZYXY0Y18} encoded the sequential and structure content of code by LSTM and tree-based LSTM and used a hybrid attention layer to obtain an integrated representation. The performance was further improved by deep reinforcement learning~\cite{ml/Williams92} to solve the exposure bias problem during the decoding phase.

Recent work on hybrid approaches ~\cite{zhangretrieval20,WeiLLXJ20}, which combine the NMT-based and IR-based methods, shows promising results. These approaches retrieve the most similar code snippets from the training set, encode the code and the retrieved result as context vectors, and decode them simultaneously. For instance, \rencos~\cite{zhangretrieval20} obtains two most similar code snippets based on the syntax-level and semantics-level information of source code, and uses them 
to generate the summary. 
\retocom

~\cite{WeiLLXJ20} retrieves the most similar code snippet and the corresponding summary, and then uses the summary as an exemplar to generate the target summary. 
\citet{haque2020improved} utilized the file context of methods (i.e. other methods in the same file) and used an attention mechanism to find words to use in summaries. However, the file context is simply modeled as token sequences of sibling methods and it lacks the class relationship information which is very important for some method summaries.

Compared with the above work, our approach also adopts the general
encoder-decoder architecture but incorporates the global contexts into
consideration with the help of GNNs, resulting in better performance on source code summarization.

\section{Conclusion}
\label{sec:con}

In this paper, we propose a new code summarization model \codeuml{}, 
which leverages two types of global context information, i.e., 
the enclosing class and class relationships, 
to assist with code summarization. 
Contemporary automatic code
summarization approaches only consider a given code snippet in the local
context,  without considering its global context information. 
We firstly represent intra-class context by using transformer-based representation of the
name of the class that a given method belongs to, to obtain the \emph{class semantic embeddings}. 

Then we use \mrgnn{} to
learn the rich structural information behind the UML class diagrams to
obtain the \emph{class relational embeddings}.

The two embeddings of class names and
UML class diagrams are incorporated into a Seq2Seq 
model
with the help of our two-level attention mechanism to 
enhance the performance of code summarization.

Experimental results have
demonstrated the effectiveness of \codeuml{} and confirmed the usefulness of
the two global context information.

We believe our work sheds some light on future research by pointing out that
source code context including local and global context can play an important
role in code summarization. In the future, we plan to explore more source code
information (e.g., call graphs, data flow graphs) 
to further improve the quality of the generated code summaries. 

\bibliographystyle{ACM-Reference-Format}
\bibliography{ref}


\begin{thebibliography}{74}


\ifx \showCODEN    \undefined \def \showCODEN     #1{\unskip}     \fi
\ifx \showDOI      \undefined \def \showDOI       #1{#1}\fi
\ifx \showISBNx    \undefined \def \showISBNx     #1{\unskip}     \fi
\ifx \showISBNxiii \undefined \def \showISBNxiii  #1{\unskip}     \fi
\ifx \showISSN     \undefined \def \showISSN      #1{\unskip}     \fi
\ifx \showLCCN     \undefined \def \showLCCN      #1{\unskip}     \fi
\ifx \shownote     \undefined \def \shownote      #1{#1}          \fi
\ifx \showarticletitle \undefined \def \showarticletitle #1{#1}   \fi
\ifx \showURL      \undefined \def \showURL       {\relax}        \fi
\providecommand\bibfield[2]{#2}
\providecommand\bibinfo[2]{#2}
\providecommand\natexlab[1]{#1}
\providecommand\showeprint[2][]{arXiv:#2}

\bibitem[\protect\citeauthoryear{Ahmad, Chakraborty, Ray, and Chang}{Ahmad
  et~al\mbox{.}}{2020}]%
        {AhmadCRC20}
\bibfield{author}{\bibinfo{person}{Wasi~Uddin Ahmad}, \bibinfo{person}{Saikat
  Chakraborty}, \bibinfo{person}{Baishakhi Ray}, {and}
  \bibinfo{person}{Kai{-}Wei Chang}.} \bibinfo{year}{2020}\natexlab{}.
\newblock \showarticletitle{A Transformer-based Approach for Source Code
  Summarization}. In \bibinfo{booktitle}{\emph{Proceedings of the 58th Annual
  Meeting of the Association for Computational Linguistics, {ACL}}}.
\newblock


\bibitem[\protect\citeauthoryear{Allamanis, Barr, Bird, and Sutton}{Allamanis
  et~al\mbox{.}}{2015a}]%
        {AllamanisBBS15}
\bibfield{author}{\bibinfo{person}{Miltiadis Allamanis},
  \bibinfo{person}{Earl~T. Barr}, \bibinfo{person}{Christian Bird}, {and}
  \bibinfo{person}{Charles~A. Sutton}.} \bibinfo{year}{2015}\natexlab{a}.
\newblock \showarticletitle{Suggesting accurate method and class names}. In
  \bibinfo{booktitle}{\emph{{ESEC/SIGSOFT} {FSE}}}. \bibinfo{pages}{38--49}.
\newblock


\bibitem[\protect\citeauthoryear{Allamanis, Peng, and Sutton}{Allamanis
  et~al\mbox{.}}{2016}]%
        {AllamanisPS16}
\bibfield{author}{\bibinfo{person}{Miltiadis Allamanis}, \bibinfo{person}{Hao
  Peng}, {and} \bibinfo{person}{Charles~A. Sutton}.}
  \bibinfo{year}{2016}\natexlab{}.
\newblock \showarticletitle{A Convolutional Attention Network for Extreme
  Summarization of Source Code}. In \bibinfo{booktitle}{\emph{{ICML}}},
  Vol.~\bibinfo{volume}{48}. \bibinfo{pages}{2091--2100}.
\newblock


\bibitem[\protect\citeauthoryear{Allamanis, Tarlow, Gordon, and Wei}{Allamanis
  et~al\mbox{.}}{2015b}]%
        {AllamanisTGW15}
\bibfield{author}{\bibinfo{person}{Miltiadis Allamanis},
  \bibinfo{person}{Daniel Tarlow}, \bibinfo{person}{Andrew~D. Gordon}, {and}
  \bibinfo{person}{Yi Wei}.} \bibinfo{year}{2015}\natexlab{b}.
\newblock \showarticletitle{Bimodal Modelling of Source Code and Natural
  Language}. In \bibinfo{booktitle}{\emph{{ICML}}}, Vol.~\bibinfo{volume}{37}.
  \bibinfo{pages}{2123--2132}.
\newblock


\bibitem[\protect\citeauthoryear{Alon, Brody, Levy, and Yahav}{Alon
  et~al\mbox{.}}{2019a}]%
        {AlonBLY19}
\bibfield{author}{\bibinfo{person}{Uri Alon}, \bibinfo{person}{Shaked Brody},
  \bibinfo{person}{Omer Levy}, {and} \bibinfo{person}{Eran Yahav}.}
  \bibinfo{year}{2019}\natexlab{a}.
\newblock \showarticletitle{code2seq: Generating Sequences from Structured
  Representations of Code}. In \bibinfo{booktitle}{\emph{{ICLR}}}.
\newblock


\bibitem[\protect\citeauthoryear{Alon, Zilberstein, Levy, and Yahav}{Alon
  et~al\mbox{.}}{2019b}]%
        {AlonZLY19}
\bibfield{author}{\bibinfo{person}{Uri Alon}, \bibinfo{person}{Meital
  Zilberstein}, \bibinfo{person}{Omer Levy}, {and} \bibinfo{person}{Eran
  Yahav}.} \bibinfo{year}{2019}\natexlab{b}.
\newblock \showarticletitle{code2vec: learning distributed representations of
  code}.
\newblock \bibinfo{journal}{\emph{{PACMPL}}} \bibinfo{volume}{3},
  \bibinfo{number}{{POPL}} (\bibinfo{year}{2019}),
  \bibinfo{pages}{40:1--40:29}.
\newblock


\bibitem[\protect\citeauthoryear{Banerjee and Lavie}{Banerjee and
  Lavie}{2005}]%
        {BanerjeeL05}
\bibfield{author}{\bibinfo{person}{Satanjeev Banerjee} {and}
  \bibinfo{person}{Alon Lavie}.} \bibinfo{year}{2005}\natexlab{}.
\newblock \showarticletitle{{METEOR:} An Automatic Metric for {MT} Evaluation
  with Improved Correlation with Human Judgments}. In
  \bibinfo{booktitle}{\emph{ACL}}. \bibinfo{publisher}{Association for
  Computational Linguistics}, \bibinfo{pages}{65--72}.
\newblock


\bibitem[\protect\citeauthoryear{Booch, Rumbaugh, and Jacobson}{Booch
  et~al\mbox{.}}{1999}]%
        {Booch99uml}
\bibfield{author}{\bibinfo{person}{Grady Booch}, \bibinfo{person}{James~E.
  Rumbaugh}, {and} \bibinfo{person}{Ivar Jacobson}.}
  \bibinfo{year}{1999}\natexlab{}.
\newblock \bibinfo{booktitle}{\emph{The unified modeling language user guide -
  the ultimate tutorial to the {UML} from the original designers}}.
\newblock \bibinfo{publisher}{Addison-Wesley-Longman}.
\newblock


\bibitem[\protect\citeauthoryear{Cer, Yang, Kong, Hua, Limtiaco, John,
  Constant, Guajardo{-}Cespedes, Yuan, Tar, Sung, Strope, and Kurzweil}{Cer
  et~al\mbox{.}}{2018}]%
        {cer2018universal}
\bibfield{author}{\bibinfo{person}{Daniel Cer}, \bibinfo{person}{Yinfei Yang},
  \bibinfo{person}{Sheng{-}yi Kong}, \bibinfo{person}{Nan Hua},
  \bibinfo{person}{Nicole Limtiaco}, \bibinfo{person}{Rhomni~St. John},
  \bibinfo{person}{Noah Constant}, \bibinfo{person}{Mario Guajardo{-}Cespedes},
  \bibinfo{person}{Steve Yuan}, \bibinfo{person}{Chris Tar},
  \bibinfo{person}{Yun{-}Hsuan Sung}, \bibinfo{person}{Brian Strope}, {and}
  \bibinfo{person}{Ray Kurzweil}.} \bibinfo{year}{2018}\natexlab{}.
\newblock \showarticletitle{Universal Sentence Encoder}.
\newblock \bibinfo{journal}{\emph{arXiv Preprint}} (\bibinfo{year}{2018}).
\newblock
\urldef\tempurl%
\url{https://arxiv.org/abs/1803.11175}
\showURL{%
\tempurl}


\bibitem[\protect\citeauthoryear{Cho, van Merrienboer, G{\"{u}}l{\c{c}}ehre,
  Bahdanau, Bougares, Schwenk, and Bengio}{Cho et~al\mbox{.}}{2014}]%
        {ChoMGBBSB14}
\bibfield{author}{\bibinfo{person}{Kyunghyun Cho}, \bibinfo{person}{Bart van
  Merrienboer}, \bibinfo{person}{{\c{C}}aglar G{\"{u}}l{\c{c}}ehre},
  \bibinfo{person}{Dzmitry Bahdanau}, \bibinfo{person}{Fethi Bougares},
  \bibinfo{person}{Holger Schwenk}, {and} \bibinfo{person}{Yoshua Bengio}.}
  \bibinfo{year}{2014}\natexlab{}.
\newblock \showarticletitle{Learning Phrase Representations using {RNN}
  Encoder-Decoder for Statistical Machine Translation}. In
  \bibinfo{booktitle}{\emph{{EMNLP}}}. \bibinfo{publisher}{{ACL}},
  \bibinfo{pages}{1724--1734}.
\newblock


\bibitem[\protect\citeauthoryear{Dam, Tran, and Pham}{Dam
  et~al\mbox{.}}{2016}]%
        {DamTP16}
\bibfield{author}{\bibinfo{person}{Hoa~Khanh Dam}, \bibinfo{person}{Truyen
  Tran}, {and} \bibinfo{person}{Trang Pham}.} \bibinfo{year}{2016}\natexlab{}.
\newblock \showarticletitle{A deep language model for software code}.
\newblock \bibinfo{journal}{\emph{arXiv Preprint}}
  \bibinfo{volume}{https://arxiv.org/abs/1608.02715} (\bibinfo{year}{2016}).
\newblock


\bibitem[\protect\citeauthoryear{Eddy, Robinson, Kraft, and Carver}{Eddy
  et~al\mbox{.}}{2013}]%
        {EddyRKC13}
\bibfield{author}{\bibinfo{person}{Brian~P. Eddy}, \bibinfo{person}{Jeffrey~A.
  Robinson}, \bibinfo{person}{Nicholas~A. Kraft}, {and}
  \bibinfo{person}{Jeffrey~C. Carver}.} \bibinfo{year}{2013}\natexlab{}.
\newblock \showarticletitle{Evaluating source code summarization techniques:
  Replication and expansion}. In \bibinfo{booktitle}{\emph{{ICPC}}}.
  \bibinfo{publisher}{{IEEE} Computer Society}, \bibinfo{pages}{13--22}.
\newblock


\bibitem[\protect\citeauthoryear{Fernandes, Allamanis, and
  Brockschmidt}{Fernandes et~al\mbox{.}}{2018}]%
        {fernandes2018structured}
\bibfield{author}{\bibinfo{person}{Patrick Fernandes},
  \bibinfo{person}{Miltiadis Allamanis}, {and} \bibinfo{person}{Marc
  Brockschmidt}.} \bibinfo{year}{2018}\natexlab{}.
\newblock \showarticletitle{Structured neural summarization}. In
  \bibinfo{booktitle}{\emph{{ICLR}}}.
\newblock


\bibitem[\protect\citeauthoryear{Fowkes, Chanthirasegaran, Ranca, Allamanis,
  Lapata, and Sutton}{Fowkes et~al\mbox{.}}{2016}]%
        {fowkes2016tassal}
\bibfield{author}{\bibinfo{person}{Jaroslav Fowkes}, \bibinfo{person}{Pankajan
  Chanthirasegaran}, \bibinfo{person}{Razvan Ranca}, \bibinfo{person}{Miltiadis
  Allamanis}, \bibinfo{person}{Mirella Lapata}, {and} \bibinfo{person}{Charles
  Sutton}.} \bibinfo{year}{2016}\natexlab{}.
\newblock \showarticletitle{TASSAL: Autofolding for source code summarization}.
  In \bibinfo{booktitle}{\emph{2016 IEEE/ACM 38th International Conference on
  Software Engineering Companion (ICSE-C)}}.
\newblock


\bibitem[\protect\citeauthoryear{Gu, Zhang, Zhang, and Kim}{Gu
  et~al\mbox{.}}{2016}]%
        {GuZZK16}
\bibfield{author}{\bibinfo{person}{Xiaodong Gu}, \bibinfo{person}{Hongyu
  Zhang}, \bibinfo{person}{Dongmei Zhang}, {and} \bibinfo{person}{Sunghun
  Kim}.} \bibinfo{year}{2016}\natexlab{}.
\newblock \showarticletitle{Deep {API} learning}. In
  \bibinfo{booktitle}{\emph{{SIGSOFT} {FSE}}}. \bibinfo{pages}{631--642}.
\newblock


\bibitem[\protect\citeauthoryear{Haiduc, Aponte, and Marcus}{Haiduc
  et~al\mbox{.}}{2010a}]%
        {HaiducAM10ICSE}
\bibfield{author}{\bibinfo{person}{Sonia Haiduc}, \bibinfo{person}{Jairo
  Aponte}, {and} \bibinfo{person}{Andrian Marcus}.}
  \bibinfo{year}{2010}\natexlab{a}.
\newblock \showarticletitle{Supporting program comprehension with source code
  summarization}. In \bibinfo{booktitle}{\emph{{ICSE}}},
  Vol.~\bibinfo{volume}{2}. \bibinfo{publisher}{{ACM}},
  \bibinfo{pages}{223--226}.
\newblock


\bibitem[\protect\citeauthoryear{Haiduc, Aponte, Moreno, and Marcus}{Haiduc
  et~al\mbox{.}}{2010b}]%
        {HaiducAMM10WCRE}
\bibfield{author}{\bibinfo{person}{Sonia Haiduc}, \bibinfo{person}{Jairo
  Aponte}, \bibinfo{person}{Laura Moreno}, {and} \bibinfo{person}{Andrian
  Marcus}.} \bibinfo{year}{2010}\natexlab{b}.
\newblock \showarticletitle{On the Use of Automated Text Summarization
  Techniques for Summarizing Source Code}. In
  \bibinfo{booktitle}{\emph{{WCRE}}}. \bibinfo{publisher}{{IEEE} Computer
  Society}, \bibinfo{pages}{35--44}.
\newblock


\bibitem[\protect\citeauthoryear{Haije}{Haije}{2016}]%
        {haije2016automatic}
\bibfield{author}{\bibinfo{person}{Tjalling Haije}.}
  \bibinfo{year}{2016}\natexlab{}.
\newblock \emph{\bibinfo{title}{Automatic comment generation using a neural
  translation model}}.
\newblock Bachelor's Thesis. \bibinfo{school}{University of Amsterdam}.
\newblock


\bibitem[\protect\citeauthoryear{Haque, LeClair, Wu, and McMillan}{Haque
  et~al\mbox{.}}{2020}]%
        {haque2020improved}
\bibfield{author}{\bibinfo{person}{Sakib Haque}, \bibinfo{person}{Alexander
  LeClair}, \bibinfo{person}{Lingfei Wu}, {and} \bibinfo{person}{Collin
  McMillan}.} \bibinfo{year}{2020}\natexlab{}.
\newblock \showarticletitle{Improved Automatic Summarization of Subroutines via
  Attention to File Context}. In \bibinfo{booktitle}{\emph{MSR}}.
\newblock


\bibitem[\protect\citeauthoryear{Hochreiter and Schmidhuber}{Hochreiter and
  Schmidhuber}{1997}]%
        {HochreiterS97}
\bibfield{author}{\bibinfo{person}{Sepp Hochreiter} {and}
  \bibinfo{person}{J{\"{u}}rgen Schmidhuber}.} \bibinfo{year}{1997}\natexlab{}.
\newblock \showarticletitle{Long Short-Term Memory}.
\newblock \bibinfo{journal}{\emph{Neural Computation}} \bibinfo{volume}{9},
  \bibinfo{number}{8} (\bibinfo{year}{1997}), \bibinfo{pages}{1735--1780}.
\newblock


\bibitem[\protect\citeauthoryear{Hu, Liu, Gomes, Zitnik, Liang, Pande, and
  Leskovec}{Hu et~al\mbox{.}}{2020}]%
        {Hu20}
\bibfield{author}{\bibinfo{person}{Weihua Hu}, \bibinfo{person}{Bowen Liu},
  \bibinfo{person}{Joseph Gomes}, \bibinfo{person}{Marinka Zitnik},
  \bibinfo{person}{Percy Liang}, \bibinfo{person}{Vijay Pande}, {and}
  \bibinfo{person}{Jure Leskovec}.} \bibinfo{year}{2020}\natexlab{}.
\newblock \showarticletitle{Strategies for Pre-training Graph Neural Networks}.
  In \bibinfo{booktitle}{\emph{ICLR}}.
\newblock


\bibitem[\protect\citeauthoryear{Hu, Li, Xia, Lo, and Jin}{Hu
  et~al\mbox{.}}{2018a}]%
        {HuLXLJ18}
\bibfield{author}{\bibinfo{person}{Xing Hu}, \bibinfo{person}{Ge Li},
  \bibinfo{person}{Xin Xia}, \bibinfo{person}{David Lo}, {and}
  \bibinfo{person}{Zhi Jin}.} \bibinfo{year}{2018}\natexlab{a}.
\newblock \showarticletitle{Deep code comment generation}. In
  \bibinfo{booktitle}{\emph{{ICPC}}}. \bibinfo{pages}{200--210}.
\newblock


\bibitem[\protect\citeauthoryear{Hu, Li, Xia, Lo, and Jin}{Hu
  et~al\mbox{.}}{2019}]%
        {hu2019deep}
\bibfield{author}{\bibinfo{person}{Xing Hu}, \bibinfo{person}{Ge Li},
  \bibinfo{person}{Xin Xia}, \bibinfo{person}{David Lo}, {and}
  \bibinfo{person}{Zhi Jin}.} \bibinfo{year}{2019}\natexlab{}.
\newblock \showarticletitle{Deep code comment generation with hybrid lexical
  and syntactical information}.
\newblock \bibinfo{journal}{\emph{Empirical Software Engineering}}
  (\bibinfo{year}{2019}).
\newblock


\bibitem[\protect\citeauthoryear{Hu, Li, Xia, Lo, Lu, and Jin}{Hu
  et~al\mbox{.}}{2018b}]%
        {HuLXLLJ18}
\bibfield{author}{\bibinfo{person}{Xing Hu}, \bibinfo{person}{Ge Li},
  \bibinfo{person}{Xin Xia}, \bibinfo{person}{David Lo}, \bibinfo{person}{Shuai
  Lu}, {and} \bibinfo{person}{Zhi Jin}.} \bibinfo{year}{2018}\natexlab{b}.
\newblock \showarticletitle{Summarizing Source Code with Transferred {API}
  Knowledge}. In \bibinfo{booktitle}{\emph{{IJCAI}}}.
  \bibinfo{publisher}{ijcai.org}, \bibinfo{pages}{2269--2275}.
\newblock


\bibitem[\protect\citeauthoryear{Husain, Wu, Gazit, Allamanis, and
  Brockschmidt}{Husain et~al\mbox{.}}{2019}]%
        {abs-1909-09436}
\bibfield{author}{\bibinfo{person}{Hamel Husain}, \bibinfo{person}{Ho{-}Hsiang
  Wu}, \bibinfo{person}{Tiferet Gazit}, \bibinfo{person}{Miltiadis Allamanis},
  {and} \bibinfo{person}{Marc Brockschmidt}.} \bibinfo{year}{2019}\natexlab{}.
\newblock \showarticletitle{CodeSearchNet Challenge: Evaluating the State of
  Semantic Code Search}.
\newblock \bibinfo{journal}{\emph{arXiv Preprint}} (\bibinfo{year}{2019}).
\newblock
\urldef\tempurl%
\url{https://arxiv.org/abs/1909.09436}
\showURL{%
\tempurl}


\bibitem[\protect\citeauthoryear{Iyer, Konstas, Cheung, and Zettlemoyer}{Iyer
  et~al\mbox{.}}{2016}]%
        {IyerKCZ16}
\bibfield{author}{\bibinfo{person}{Srinivasan Iyer}, \bibinfo{person}{Ioannis
  Konstas}, \bibinfo{person}{Alvin Cheung}, {and} \bibinfo{person}{Luke
  Zettlemoyer}.} \bibinfo{year}{2016}\natexlab{}.
\newblock \showarticletitle{Summarizing Source Code using a Neural Attention
  Model}. In \bibinfo{booktitle}{\emph{{ACL}}}, Vol.~\bibinfo{volume}{1}.
  \bibinfo{publisher}{The Association for Computer Linguistics}.
\newblock


\bibitem[\protect\citeauthoryear{Kipf and Welling}{Kipf and Welling}{2017}]%
        {KipfW17}
\bibfield{author}{\bibinfo{person}{Thomas~N. Kipf} {and} \bibinfo{person}{Max
  Welling}.} \bibinfo{year}{2017}\natexlab{}.
\newblock \showarticletitle{Semi-Supervised Classification with Graph
  Convolutional Networks}. In \bibinfo{booktitle}{\emph{{ICLR}}}.
\newblock


\bibitem[\protect\citeauthoryear{LeClair, Haque, Wu, and McMillan}{LeClair
  et~al\mbox{.}}{2020}]%
        {leclair2020gnn}
\bibfield{author}{\bibinfo{person}{Alexander LeClair}, \bibinfo{person}{Sakib
  Haque}, \bibinfo{person}{Linfgei Wu}, {and} \bibinfo{person}{Collin
  McMillan}.} \bibinfo{year}{2020}\natexlab{}.
\newblock \showarticletitle{Improved code summarization via a graph neural
  network}. In \bibinfo{booktitle}{\emph{{ICPC}}}.
\newblock


\bibitem[\protect\citeauthoryear{LeClair, Jiang, and McMillan}{LeClair
  et~al\mbox{.}}{2019}]%
        {LeClairJM19}
\bibfield{author}{\bibinfo{person}{Alexander LeClair}, \bibinfo{person}{Siyuan
  Jiang}, {and} \bibinfo{person}{Collin McMillan}.}
  \bibinfo{year}{2019}\natexlab{}.
\newblock \showarticletitle{A neural model for generating natural language
  summaries of program subroutines}. In \bibinfo{booktitle}{\emph{{ICSE}}}.
  \bibinfo{pages}{795--806}.
\newblock


\bibitem[\protect\citeauthoryear{LeCun, Bengio, and Hinton}{LeCun
  et~al\mbox{.}}{2015}]%
        {lecun2015deep}
\bibfield{author}{\bibinfo{person}{Yann LeCun}, \bibinfo{person}{Yoshua
  Bengio}, {and} \bibinfo{person}{Geoffrey Hinton}.}
  \bibinfo{year}{2015}\natexlab{}.
\newblock \showarticletitle{Deep Learning}.
\newblock \bibinfo{journal}{\emph{Nature}} \bibinfo{volume}{521},
  \bibinfo{number}{7553} (\bibinfo{year}{2015}), \bibinfo{pages}{436--444}.
\newblock


\bibitem[\protect\citeauthoryear{Li, Wang, Nguyen, and Van~Nguyen}{Li
  et~al\mbox{.}}{2019}]%
        {li2019improving}
\bibfield{author}{\bibinfo{person}{Yi Li}, \bibinfo{person}{Shaohua Wang},
  \bibinfo{person}{Tien~N Nguyen}, {and} \bibinfo{person}{Son Van~Nguyen}.}
  \bibinfo{year}{2019}\natexlab{}.
\newblock \showarticletitle{Improving bug detection via context-based code
  representation learning and attention-based neural networks}. In
  \bibinfo{booktitle}{\emph{OOPSLA}}.
\newblock


\bibitem[\protect\citeauthoryear{Lin}{Lin}{2004}]%
        {lin-2004-rouge}
\bibfield{author}{\bibinfo{person}{Chin{-}Yew Lin}.}
  \bibinfo{year}{2004}\natexlab{}.
\newblock \showarticletitle{{ROUGE}: A Package for Automatic Evaluation of
  Summaries}. In \bibinfo{booktitle}{\emph{Text Summarization Branches Out}}.
  \bibinfo{pages}{74--81}.
\newblock


\bibitem[\protect\citeauthoryear{Lin and Och}{Lin and Och}{2004}]%
        {LinO04}
\bibfield{author}{\bibinfo{person}{Chin{-}Yew Lin} {and}
  \bibinfo{person}{Franz~Josef Och}.} \bibinfo{year}{2004}\natexlab{}.
\newblock \showarticletitle{{ORANGE:} a Method for Evaluating Automatic
  Evaluation Metrics for Machine Translation}. In
  \bibinfo{booktitle}{\emph{{COLING}}}.
\newblock


\bibitem[\protect\citeauthoryear{Ling, Blunsom, Grefenstette, Hermann,
  Kocisk{\'{y}}, Wang, and Senior}{Ling et~al\mbox{.}}{2016}]%
        {LingBGHKWS16}
\bibfield{author}{\bibinfo{person}{Wang Ling}, \bibinfo{person}{Phil Blunsom},
  \bibinfo{person}{Edward Grefenstette}, \bibinfo{person}{Karl~Moritz Hermann},
  \bibinfo{person}{Tom{\'{a}}s Kocisk{\'{y}}}, \bibinfo{person}{Fumin Wang},
  {and} \bibinfo{person}{Andrew~W. Senior}.} \bibinfo{year}{2016}\natexlab{}.
\newblock \showarticletitle{Latent Predictor Networks for Code Generation}. In
  \bibinfo{booktitle}{\emph{{ACL}}}, Vol.~\bibinfo{volume}{1}.
  \bibinfo{publisher}{The Association for Computer Linguistics}.
\newblock


\bibitem[\protect\citeauthoryear{Liu, Xia, Hassan, Lo, Xing, and Wang}{Liu
  et~al\mbox{.}}{2018}]%
        {LiuXHLXW18}
\bibfield{author}{\bibinfo{person}{Zhongxin Liu}, \bibinfo{person}{Xin Xia},
  \bibinfo{person}{Ahmed~E. Hassan}, \bibinfo{person}{David Lo},
  \bibinfo{person}{Zhenchang Xing}, {and} \bibinfo{person}{Xinyu Wang}.}
  \bibinfo{year}{2018}\natexlab{}.
\newblock \showarticletitle{Neural-machine-translation-based commit message
  generation: how far are we?}. In \bibinfo{booktitle}{\emph{{ASE}}}.
\newblock


\bibitem[\protect\citeauthoryear{Liu, Xia, Treude, Lo, and Li}{Liu
  et~al\mbox{.}}{2019}]%
        {Liu0T0L19}
\bibfield{author}{\bibinfo{person}{Zhongxin Liu}, \bibinfo{person}{Xin Xia},
  \bibinfo{person}{Christoph Treude}, \bibinfo{person}{David Lo}, {and}
  \bibinfo{person}{Shanping Li}.} \bibinfo{year}{2019}\natexlab{}.
\newblock \showarticletitle{Automatic Generation of Pull Request Descriptions}.
  In \bibinfo{booktitle}{\emph{{ASE}}}.
\newblock


\bibitem[\protect\citeauthoryear{Loshchilov and Hutter}{Loshchilov and
  Hutter}{2019}]%
        {loshchilov2017decoupled}
\bibfield{author}{\bibinfo{person}{Ilya Loshchilov} {and}
  \bibinfo{person}{Frank Hutter}.} \bibinfo{year}{2019}\natexlab{}.
\newblock \showarticletitle{Decoupled Weight Decay Regularization}. In
  \bibinfo{booktitle}{\emph{{ICLR}}}.
\newblock


\bibitem[\protect\citeauthoryear{Luong, Pham, and Manning}{Luong
  et~al\mbox{.}}{2015}]%
        {LuongPM15}
\bibfield{author}{\bibinfo{person}{Thang Luong}, \bibinfo{person}{Hieu Pham},
  {and} \bibinfo{person}{Christopher~D. Manning}.}
  \bibinfo{year}{2015}\natexlab{}.
\newblock \showarticletitle{Effective Approaches to Attention-based Neural
  Machine Translation}. In \bibinfo{booktitle}{\emph{{EMNLP}}}.
  \bibinfo{pages}{1412--1421}.
\newblock


\bibitem[\protect\citeauthoryear{McBurney and McMillan}{McBurney and
  McMillan}{2014}]%
        {mcburney2014automatic}
\bibfield{author}{\bibinfo{person}{Paul~W McBurney} {and}
  \bibinfo{person}{Collin McMillan}.} \bibinfo{year}{2014}\natexlab{}.
\newblock \showarticletitle{Automatic documentation generation via source code
  summarization of method context}. In \bibinfo{booktitle}{\emph{{ICPC}}}.
\newblock


\bibitem[\protect\citeauthoryear{Mou, Li, Zhang, Wang, and Jin}{Mou
  et~al\mbox{.}}{2016}]%
        {MouLZWJ16}
\bibfield{author}{\bibinfo{person}{Lili Mou}, \bibinfo{person}{Ge Li},
  \bibinfo{person}{Lu Zhang}, \bibinfo{person}{Tao Wang}, {and}
  \bibinfo{person}{Zhi Jin}.} \bibinfo{year}{2016}\natexlab{}.
\newblock \showarticletitle{Convolutional Neural Networks over Tree Structures
  for Programming Language Processing}. In \bibinfo{booktitle}{\emph{{AAAI}}}.
  \bibinfo{pages}{1287--1293}.
\newblock


\bibitem[\protect\citeauthoryear{Movshovitz{-}Attias and
  Cohen}{Movshovitz{-}Attias and Cohen}{2013}]%
        {Movshovitz-AttiasC13}
\bibfield{author}{\bibinfo{person}{Dana Movshovitz{-}Attias} {and}
  \bibinfo{person}{William~W. Cohen}.} \bibinfo{year}{2013}\natexlab{}.
\newblock \showarticletitle{Natural Language Models for Predicting Programming
  Comments}. In \bibinfo{booktitle}{\emph{{ACL}}}, Vol.~\bibinfo{volume}{2}.
  \bibinfo{pages}{35--40}.
\newblock


\bibitem[\protect\citeauthoryear{Oracle}{Oracle}{2018}]%
        {javadoc}
\bibfield{author}{\bibinfo{person}{Oracle}.} \bibinfo{year}{2018}\natexlab{}.
\newblock \showarticletitle{How to write doc comments for the javadoc tool}. In
  \bibinfo{booktitle}{\emph{\url{https://www.
  oracle.com/technetwork/articles/java/index-137868.html}}}.
\newblock


\bibitem[\protect\citeauthoryear{Papineni, Roukos, Ward, and Zhu}{Papineni
  et~al\mbox{.}}{2002}]%
        {PapineniRWZ02}
\bibfield{author}{\bibinfo{person}{Kishore Papineni}, \bibinfo{person}{Salim
  Roukos}, \bibinfo{person}{Todd Ward}, {and} \bibinfo{person}{Wei{-}Jing
  Zhu}.} \bibinfo{year}{2002}\natexlab{}.
\newblock \showarticletitle{Bleu: a Method for Automatic Evaluation of Machine
  Translation}. In \bibinfo{booktitle}{\emph{{ACL}}}.
  \bibinfo{publisher}{{ACL}}, \bibinfo{pages}{311--318}.
\newblock


\bibitem[\protect\citeauthoryear{Parisotto, Mohamed, Singh, Li, Zhou, and
  Kohli}{Parisotto et~al\mbox{.}}{2016}]%
        {ParisottoMSLZK16}
\bibfield{author}{\bibinfo{person}{Emilio Parisotto},
  \bibinfo{person}{Abdel{-}rahman Mohamed}, \bibinfo{person}{Rishabh Singh},
  \bibinfo{person}{Lihong Li}, \bibinfo{person}{Dengyong Zhou}, {and}
  \bibinfo{person}{Pushmeet Kohli}.} \bibinfo{year}{2016}\natexlab{}.
\newblock \showarticletitle{Neuro-Symbolic Program Synthesis}.
\newblock \bibinfo{journal}{\emph{arXiv Preprint}}
  \bibinfo{volume}{https://arxiv.org/abs/1611.01855} (\bibinfo{year}{2016}).
\newblock


\bibitem[\protect\citeauthoryear{Piech, Huang, Nguyen, Phulsuksombati, Sahami,
  and Guibas}{Piech et~al\mbox{.}}{2015}]%
        {PiechHNPSG15}
\bibfield{author}{\bibinfo{person}{Chris Piech}, \bibinfo{person}{Jonathan
  Huang}, \bibinfo{person}{Andy Nguyen}, \bibinfo{person}{Mike Phulsuksombati},
  \bibinfo{person}{Mehran Sahami}, {and} \bibinfo{person}{Leonidas~J. Guibas}.}
  \bibinfo{year}{2015}\natexlab{}.
\newblock \showarticletitle{Learning Program Embeddings to Propagate Feedback
  on Student Code}. In \bibinfo{booktitle}{\emph{{ICML}}},
  Vol.~\bibinfo{volume}{37}. \bibinfo{pages}{1093--1102}.
\newblock


\bibitem[\protect\citeauthoryear{Rodeghero, McMillan, McBurney, Bosch, and
  D'Mello}{Rodeghero et~al\mbox{.}}{2014}]%
        {RodegheroMMBD14}
\bibfield{author}{\bibinfo{person}{Paige Rodeghero}, \bibinfo{person}{Collin
  McMillan}, \bibinfo{person}{Paul~W. McBurney}, \bibinfo{person}{Nigel Bosch},
  {and} \bibinfo{person}{Sidney~K. D'Mello}.} \bibinfo{year}{2014}\natexlab{}.
\newblock \showarticletitle{Improving automated source code summarization via
  an eye-tracking study of programmers}. In \bibinfo{booktitle}{\emph{{ICSE}}}.
  \bibinfo{publisher}{{ACM}}, \bibinfo{pages}{390--401}.
\newblock


\bibitem[\protect\citeauthoryear{Rumbaugh, Jacobson, and Booch}{Rumbaugh
  et~al\mbox{.}}{1999}]%
        {Rumbaugh99uml}
\bibfield{author}{\bibinfo{person}{James~E. Rumbaugh}, \bibinfo{person}{Ivar
  Jacobson}, {and} \bibinfo{person}{Grady Booch}.}
  \bibinfo{year}{1999}\natexlab{}.
\newblock \bibinfo{booktitle}{\emph{The unified modeling language reference
  manual}}.
\newblock \bibinfo{publisher}{Addison-Wesley-Longman}.
\newblock


\bibitem[\protect\citeauthoryear{Rumelhart, Hinton, and Williams}{Rumelhart
  et~al\mbox{.}}{1986}]%
        {rumelhart1986learning}
\bibfield{author}{\bibinfo{person}{David~E Rumelhart},
  \bibinfo{person}{Geoffrey~E Hinton}, {and} \bibinfo{person}{Ronald~J
  Williams}.} \bibinfo{year}{1986}\natexlab{}.
\newblock \showarticletitle{Learning representations by back-propagating
  errors}.
\newblock \bibinfo{journal}{\emph{Nature}} \bibinfo{volume}{323},
  \bibinfo{number}{6088} (\bibinfo{year}{1986}), \bibinfo{pages}{533--536}.
\newblock


\bibitem[\protect\citeauthoryear{Rush, Chopra, and Weston}{Rush
  et~al\mbox{.}}{2015}]%
        {RushCW15}
\bibfield{author}{\bibinfo{person}{Alexander~M. Rush}, \bibinfo{person}{Sumit
  Chopra}, {and} \bibinfo{person}{Jason Weston}.}
  \bibinfo{year}{2015}\natexlab{}.
\newblock \showarticletitle{A Neural Attention Model for Abstractive Sentence
  Summarization}. In \bibinfo{booktitle}{\emph{{EMNLP}}}.
  \bibinfo{pages}{379--389}.
\newblock


\bibitem[\protect\citeauthoryear{Sajnani, Saini, Svajlenko, Roy, and
  Lopes}{Sajnani et~al\mbox{.}}{2016}]%
        {SajnaniSSRL16}
\bibfield{author}{\bibinfo{person}{Hitesh Sajnani}, \bibinfo{person}{Vaibhav
  Saini}, \bibinfo{person}{Jeffrey Svajlenko}, \bibinfo{person}{Chanchal~K.
  Roy}, {and} \bibinfo{person}{Cristina~V. Lopes}.}
  \bibinfo{year}{2016}\natexlab{}.
\newblock \showarticletitle{SourcererCC: scaling code clone detection to
  big-code}. In \bibinfo{booktitle}{\emph{{ICSE}}}.
\newblock


\bibitem[\protect\citeauthoryear{See, Liu, and Manning}{See
  et~al\mbox{.}}{2017}]%
        {SeeLM17}
\bibfield{author}{\bibinfo{person}{Abigail See}, \bibinfo{person}{Peter~J.
  Liu}, {and} \bibinfo{person}{Christopher~D. Manning}.}
  \bibinfo{year}{2017}\natexlab{}.
\newblock \showarticletitle{Get To The Point: Summarization with
  Pointer-Generator Networks}. In \bibinfo{booktitle}{\emph{{ACL}}}.
\newblock


\bibitem[\protect\citeauthoryear{Spinellis}{Spinellis}{2010}]%
        {spinellis2010code}
\bibfield{author}{\bibinfo{person}{Diomidis Spinellis}.}
  \bibinfo{year}{2010}\natexlab{}.
\newblock \showarticletitle{Code documentation}.
\newblock \bibinfo{journal}{\emph{IEEE software}} \bibinfo{volume}{27},
  \bibinfo{number}{4} (\bibinfo{year}{2010}), \bibinfo{pages}{18--19}.
\newblock


\bibitem[\protect\citeauthoryear{Sridhara, Hill, Muppaneni, Pollock, and
  Vijay{-}Shanker}{Sridhara et~al\mbox{.}}{2010}]%
        {SridharaHMPV10}
\bibfield{author}{\bibinfo{person}{Giriprasad Sridhara}, \bibinfo{person}{Emily
  Hill}, \bibinfo{person}{Divya Muppaneni}, \bibinfo{person}{Lori~L. Pollock},
  {and} \bibinfo{person}{K. Vijay{-}Shanker}.} \bibinfo{year}{2010}\natexlab{}.
\newblock \showarticletitle{Towards automatically generating summary comments
  for Java methods}. In \bibinfo{booktitle}{\emph{{ASE}}}.
  \bibinfo{pages}{43--52}.
\newblock


\bibitem[\protect\citeauthoryear{Sui, Cheng, Zhang, and Wang}{Sui
  et~al\mbox{.}}{2020}]%
        {SuiCZ020}
\bibfield{author}{\bibinfo{person}{Yulei Sui}, \bibinfo{person}{Xiao Cheng},
  \bibinfo{person}{Guanqin Zhang}, {and} \bibinfo{person}{Haoyu Wang}.}
  \bibinfo{year}{2020}\natexlab{}.
\newblock \showarticletitle{Flow2Vec: value-flow-based precise code embedding}.
  In \bibinfo{booktitle}{\emph{{OOPSLA}}}.
\newblock


\bibitem[\protect\citeauthoryear{Sutskever, Vinyals, and Le}{Sutskever
  et~al\mbox{.}}{2014}]%
        {SutskeverVL14}
\bibfield{author}{\bibinfo{person}{Ilya Sutskever}, \bibinfo{person}{Oriol
  Vinyals}, {and} \bibinfo{person}{Quoc~V. Le}.}
  \bibinfo{year}{2014}\natexlab{}.
\newblock \showarticletitle{Sequence to Sequence Learning with Neural
  Networks}. In \bibinfo{booktitle}{\emph{{NIPS}}}.
  \bibinfo{pages}{3104--3112}.
\newblock


\bibitem[\protect\citeauthoryear{Vaswani, Shazeer, Parmar, Uszkoreit, Jones,
  Gomez, Kaiser, and Polosukhin}{Vaswani et~al\mbox{.}}{2017}]%
        {VaswaniSPUJGKP17}
\bibfield{author}{\bibinfo{person}{Ashish Vaswani}, \bibinfo{person}{Noam
  Shazeer}, \bibinfo{person}{Niki Parmar}, \bibinfo{person}{Jakob Uszkoreit},
  \bibinfo{person}{Llion Jones}, \bibinfo{person}{Aidan~N. Gomez},
  \bibinfo{person}{Lukasz Kaiser}, {and} \bibinfo{person}{Illia Polosukhin}.}
  \bibinfo{year}{2017}\natexlab{}.
\newblock \showarticletitle{Attention is All you Need}. In
  \bibinfo{booktitle}{\emph{{NIPS}}}. \bibinfo{pages}{5998--6008}.
\newblock


\bibitem[\protect\citeauthoryear{Vedantam, Zitnick, and Parikh}{Vedantam
  et~al\mbox{.}}{2015}]%
        {VedantamZP15}
\bibfield{author}{\bibinfo{person}{Ramakrishna Vedantam},
  \bibinfo{person}{C.~Lawrence Zitnick}, {and} \bibinfo{person}{Devi Parikh}.}
  \bibinfo{year}{2015}\natexlab{}.
\newblock \showarticletitle{CIDEr: Consensus-based image description
  evaluation}. In \bibinfo{booktitle}{\emph{{CVPR}}}.
  \bibinfo{publisher}{{IEEE} Computer Society}, \bibinfo{pages}{4566--4575}.
\newblock


\bibitem[\protect\citeauthoryear{Velickovic, Cucurull, Casanova, Romero,
  Li{\`{o}}, and Bengio}{Velickovic et~al\mbox{.}}{2018}]%
        {VelickovicCCRLB18}
\bibfield{author}{\bibinfo{person}{Petar Velickovic}, \bibinfo{person}{Guillem
  Cucurull}, \bibinfo{person}{Arantxa Casanova}, \bibinfo{person}{Adriana
  Romero}, \bibinfo{person}{Pietro Li{\`{o}}}, {and} \bibinfo{person}{Yoshua
  Bengio}.} \bibinfo{year}{2018}\natexlab{}.
\newblock \showarticletitle{Graph Attention Networks}. In
  \bibinfo{booktitle}{\emph{{ICLR}}}.
\newblock


\bibitem[\protect\citeauthoryear{Vinyals, Toshev, Bengio, and Erhan}{Vinyals
  et~al\mbox{.}}{2015}]%
        {VinyalsTBE15}
\bibfield{author}{\bibinfo{person}{Oriol Vinyals}, \bibinfo{person}{Alexander
  Toshev}, \bibinfo{person}{Samy Bengio}, {and} \bibinfo{person}{Dumitru
  Erhan}.} \bibinfo{year}{2015}\natexlab{}.
\newblock \showarticletitle{Show and tell: {A} neural image caption generator}.
  In \bibinfo{booktitle}{\emph{{CVPR}}}. \bibinfo{pages}{3156--3164}.
\newblock


\bibitem[\protect\citeauthoryear{Wan, Zhao, Yang, Xu, Ying, Wu, and Yu}{Wan
  et~al\mbox{.}}{2018}]%
        {WanZYXY0Y18}
\bibfield{author}{\bibinfo{person}{Yao Wan}, \bibinfo{person}{Zhou Zhao},
  \bibinfo{person}{Min Yang}, \bibinfo{person}{Guandong Xu},
  \bibinfo{person}{Haochao Ying}, \bibinfo{person}{Jian Wu}, {and}
  \bibinfo{person}{Philip~S. Yu}.} \bibinfo{year}{2018}\natexlab{}.
\newblock \showarticletitle{Improving automatic source code summarization via
  deep reinforcement learning}. In \bibinfo{booktitle}{\emph{{ASE}}}.
  \bibinfo{publisher}{{ACM}}, \bibinfo{pages}{397--407}.
\newblock


\bibitem[\protect\citeauthoryear{Wei, Li, Xia, Fu, and Jin}{Wei
  et~al\mbox{.}}{2019}]%
        {wei2019code}
\bibfield{author}{\bibinfo{person}{Bolin Wei}, \bibinfo{person}{Ge Li},
  \bibinfo{person}{Xin Xia}, \bibinfo{person}{Zhiyi Fu}, {and}
  \bibinfo{person}{Zhi Jin}.} \bibinfo{year}{2019}\natexlab{}.
\newblock \showarticletitle{Code generation as a dual task of code
  summarization}. In \bibinfo{booktitle}{\emph{Advances in Neural Information
  Processing Systems}}.
\newblock


\bibitem[\protect\citeauthoryear{Wei, Li, Li, Xia, and Jin}{Wei
  et~al\mbox{.}}{2020}]%
        {WeiLLXJ20}
\bibfield{author}{\bibinfo{person}{Bolin Wei}, \bibinfo{person}{Yongmin Li},
  \bibinfo{person}{Ge Li}, \bibinfo{person}{Xin Xia}, {and}
  \bibinfo{person}{Zhi Jin}.} \bibinfo{year}{2020}\natexlab{}.
\newblock \showarticletitle{Retrieve and Refine: Exemplar-based Neural Comment
  Generation}. In \bibinfo{booktitle}{\emph{{ASE}}}.
  \bibinfo{publisher}{{IEEE}}, \bibinfo{pages}{349--360}.
\newblock


\bibitem[\protect\citeauthoryear{Wilcoxon, Katti, and Wilcox}{Wilcoxon
  et~al\mbox{.}}{1970}]%
        {wilcoxon1970critical}
\bibfield{author}{\bibinfo{person}{Frank Wilcoxon}, \bibinfo{person}{SK Katti},
  {and} \bibinfo{person}{Roberta~A Wilcox}.} \bibinfo{year}{1970}\natexlab{}.
\newblock \showarticletitle{Critical values and probability levels for the
  Wilcoxon rank sum test and the Wilcoxon signed rank test}.
\newblock \bibinfo{journal}{\emph{Selected tables in mathematical statistics}}
  \bibinfo{volume}{1} (\bibinfo{year}{1970}), \bibinfo{pages}{171--259}.
\newblock


\bibitem[\protect\citeauthoryear{Williams}{Williams}{1992}]%
        {ml/Williams92}
\bibfield{author}{\bibinfo{person}{Ronald~J. Williams}.}
  \bibinfo{year}{1992}\natexlab{}.
\newblock \showarticletitle{Simple Statistical Gradient-Following Algorithms
  for Connectionist Reinforcement Learning}.
\newblock \bibinfo{journal}{\emph{Mach. Learn.}}  \bibinfo{volume}{8}
  (\bibinfo{year}{1992}), \bibinfo{pages}{229--256}.
\newblock


\bibitem[\protect\citeauthoryear{Williams and Zipser}{Williams and
  Zipser}{1989}]%
        {WilliamsZ89}
\bibfield{author}{\bibinfo{person}{Ronald~J. Williams} {and}
  \bibinfo{person}{David Zipser}.} \bibinfo{year}{1989}\natexlab{}.
\newblock \showarticletitle{A Learning Algorithm for Continually Running Fully
  Recurrent Neural Networks}.
\newblock \bibinfo{journal}{\emph{Neural Computation}} \bibinfo{volume}{1},
  \bibinfo{number}{2} (\bibinfo{year}{1989}), \bibinfo{pages}{270--280}.
\newblock


\bibitem[\protect\citeauthoryear{Wong, Liu, and Tan}{Wong
  et~al\mbox{.}}{2015}]%
        {WongLT15}
\bibfield{author}{\bibinfo{person}{Edmund Wong}, \bibinfo{person}{Taiyue Liu},
  {and} \bibinfo{person}{Lin Tan}.} \bibinfo{year}{2015}\natexlab{}.
\newblock \showarticletitle{CloCom: Mining existing source code for automatic
  comment generation}. In \bibinfo{booktitle}{\emph{{SANER}}}.
  \bibinfo{publisher}{{IEEE} Computer Society}, \bibinfo{pages}{380--389}.
\newblock


\bibitem[\protect\citeauthoryear{Wong, Yang, and Tan}{Wong
  et~al\mbox{.}}{2013}]%
        {WongYT13}
\bibfield{author}{\bibinfo{person}{Edmund Wong}, \bibinfo{person}{Jinqiu Yang},
  {and} \bibinfo{person}{Lin Tan}.} \bibinfo{year}{2013}\natexlab{}.
\newblock \showarticletitle{AutoComment: Mining question and answer sites for
  automatic comment generation}. In \bibinfo{booktitle}{\emph{{ASE}}}.
  \bibinfo{publisher}{{IEEE}}, \bibinfo{pages}{562--567}.
\newblock


\bibitem[\protect\citeauthoryear{Wu, Zhang, Gao, He, Weng, Gao, and Chen}{Wu
  et~al\mbox{.}}{2019}]%
        {WuZGHWGC19}
\bibfield{author}{\bibinfo{person}{Qitian Wu}, \bibinfo{person}{Hengrui Zhang},
  \bibinfo{person}{Xiaofeng Gao}, \bibinfo{person}{Peng He},
  \bibinfo{person}{Paul Weng}, \bibinfo{person}{Han Gao}, {and}
  \bibinfo{person}{Guihai Chen}.} \bibinfo{year}{2019}\natexlab{}.
\newblock \showarticletitle{Dual Graph Attention Networks for Deep Latent
  Representation of Multifaceted Social Effects in Recommender Systems}. In
  \bibinfo{booktitle}{\emph{{WWW}}}. \bibinfo{pages}{2091--2102}.
\newblock


\bibitem[\protect\citeauthoryear{Wu, Pan, Chen, Long, Zhang, and Yu}{Wu
  et~al\mbox{.}}{2020}]%
        {WuPCLZY20}
\bibfield{author}{\bibinfo{person}{Zonghan Wu}, \bibinfo{person}{Shirui Pan},
  \bibinfo{person}{Fengwen Chen}, \bibinfo{person}{Guodong Long},
  \bibinfo{person}{Chengqi Zhang}, {and} \bibinfo{person}{Philip~S. Yu}.}
  \bibinfo{year}{2020}\natexlab{}.
\newblock \showarticletitle{A Comprehensive Survey on Graph Neural Networks}.
\newblock \bibinfo{journal}{\emph{{IEEE} Trans. Knowl. Data Eng.}}
  (\bibinfo{year}{2020}).
\newblock


\bibitem[\protect\citeauthoryear{Young, Hazarika, Poria, and Cambria}{Young
  et~al\mbox{.}}{2018}]%
        {YoungHPC18}
\bibfield{author}{\bibinfo{person}{Tom Young}, \bibinfo{person}{Devamanyu
  Hazarika}, \bibinfo{person}{Soujanya Poria}, {and} \bibinfo{person}{Erik
  Cambria}.} \bibinfo{year}{2018}\natexlab{}.
\newblock \showarticletitle{Recent Trends in Deep Learning Based Natural
  Language Processing [Review Article]}.
\newblock \bibinfo{journal}{\emph{{IEEE} Comput. Intell. Mag.}}
  \bibinfo{volume}{13}, \bibinfo{number}{3} (\bibinfo{year}{2018}),
  \bibinfo{pages}{55--75}.
\newblock


\bibitem[\protect\citeauthoryear{Zhang, Shi, Xie, Ma, King, and Yeung}{Zhang
  et~al\mbox{.}}{2018}]%
        {ZhangSXMKY18}
\bibfield{author}{\bibinfo{person}{Jiani Zhang}, \bibinfo{person}{Xingjian
  Shi}, \bibinfo{person}{Junyuan Xie}, \bibinfo{person}{Hao Ma},
  \bibinfo{person}{Irwin King}, {and} \bibinfo{person}{Dit{-}Yan Yeung}.}
  \bibinfo{year}{2018}\natexlab{}.
\newblock \showarticletitle{GaAN: Gated Attention Networks for Learning on
  Large and Spatiotemporal Graphs}. In \bibinfo{booktitle}{\emph{{UAI}}}.
  \bibinfo{pages}{339--349}.
\newblock


\bibitem[\protect\citeauthoryear{Zhang, Wang, Zhang, Sun, and Liu}{Zhang
  et~al\mbox{.}}{2020}]%
        {zhangretrieval20}
\bibfield{author}{\bibinfo{person}{Jian Zhang}, \bibinfo{person}{Xu Wang},
  \bibinfo{person}{Hongyu Zhang}, \bibinfo{person}{Hailong Sun}, {and}
  \bibinfo{person}{Xudong Liu}.} \bibinfo{year}{2020}\natexlab{}.
\newblock \showarticletitle{Retrieval-based Neural Source Code Summarization}.
  In \bibinfo{booktitle}{\emph{{ICSE}}}. \bibinfo{publisher}{{IEEE} / {ACM}}.
\newblock


\bibitem[\protect\citeauthoryear{Zhang, Wang, Zhang, Sun, Wang, and Liu}{Zhang
  et~al\mbox{.}}{2019}]%
        {ZhangWZ0WL19}
\bibfield{author}{\bibinfo{person}{Jian Zhang}, \bibinfo{person}{Xu Wang},
  \bibinfo{person}{Hongyu Zhang}, \bibinfo{person}{Hailong Sun},
  \bibinfo{person}{Kaixuan Wang}, {and} \bibinfo{person}{Xudong Liu}.}
  \bibinfo{year}{2019}\natexlab{}.
\newblock \showarticletitle{A novel neural source code representation based on
  abstract syntax tree}. In \bibinfo{booktitle}{\emph{{ICSE}}}.
  \bibinfo{pages}{783--794}.
\newblock


\bibitem[\protect\citeauthoryear{Zhou, Cui, Zhang, Yang, Liu, and Sun}{Zhou
  et~al\mbox{.}}{2018}]%
        {abs-1812-08434}
\bibfield{author}{\bibinfo{person}{Jie Zhou}, \bibinfo{person}{Ganqu Cui},
  \bibinfo{person}{Zhengyan Zhang}, \bibinfo{person}{Cheng Yang},
  \bibinfo{person}{Zhiyuan Liu}, {and} \bibinfo{person}{Maosong Sun}.}
  \bibinfo{year}{2018}\natexlab{}.
\newblock \showarticletitle{Graph Neural Networks: {A} Review of Methods and
  Applications}.
\newblock \bibinfo{journal}{\emph{arXiv Preprint}} (\bibinfo{year}{2018}).
\newblock
\urldef\tempurl%
\url{https://arxiv.org/abs/1812.08434}
\showURL{%
\tempurl}


\end{thebibliography}

\end{document}